\pdfoutput=1

\documentclass[11pt,twoside,a4paper,cmspaper,final,collab]{cms-tdr}
\def\svnVersion{467299P}\def\cmsCernNoTag{CERN-EP-2018-027}\def\cmsCernDate{\today}\def\cmsMessage{Published in the Journal of High Energy Physics as \href{http://dx.doi.org/10.1007/JHEP06(2018)120}{\doi{10.1007/JHEP06(2018)120.}}}

\begin{document}\cmsNoteHeader{EXO-16-047}

\hyphenation{had-ron-i-za-tion}
\hyphenation{cal-or-i-me-ter}
\hyphenation{de-vices}
\RCS$HeadURL: svn+ssh://svn.cern.ch/reps/tdr2/papers/EXO-16-047/trunk/EXO-16-047.tex $
\RCS$Id: EXO-16-047.tex 467090 2018-07-02 15:41:01Z gdaskal $

\newlength\cmsFigWidth
\ifthenelse{\boolean{cms@external}}{\setlength\cmsFigWidth{0.85\columnwidth}}{\setlength\cmsFigWidth{0.4\textwidth}}
\ifthenelse{\boolean{cms@external}}{\providecommand{\cmsLeft}{top\xspace}}{\providecommand{\cmsLeft}{left\xspace}}
\ifthenelse{\boolean{cms@external}}{\providecommand{\cmsRight}{bottom\xspace}}{\providecommand{\cmsRight}{right\xspace}}
\newcommand{\anaLumiee}{35.9}
\newcommand{\anaLumimumu}{36.3}
\newcommand{\limitZssm}{4.50}
\newcommand{\limitZpsi}{3.90}
\newcommand{\limitGone}{2.10}
\newcommand{\limitGtwo}{3.65}
\newcommand{\limitGthree}{4.25}
\newcommand{\cPZg}{\ensuremath{\cmsSymbolFace{Z}/\gamma^{*}}\xspace}
\newcommand{\GKK}{\ensuremath{\mathrm{G}_\mathrm{KK}}\xspace}
\newcommand{\ZPSSM}{\ensuremath{\cPZpr_\mathrm{SSM}}\xspace}
\newcommand{\ZPPSI}{\ensuremath{\cPZpr_\psi}\xspace}

\cmsNoteHeader{EXO-16-047}
\title{
Search for high-mass resonances in dilepton final states in proton-proton collisions at $\sqrt{s}=13\TeV$
}

\date{\today}

\abstract{
A search is presented for new high-mass resonances decaying into electron or muon pairs. The search uses proton-proton collision data at a centre-of-mass energy of 13\TeV collected by the CMS experiment at the LHC in 2016, corresponding to an integrated luminosity of 36\fbinv. Observations are in agreement with standard model expectations. Upper limits on the product of a new resonance production cross section and branching fraction to dileptons are calculated in a model-independent manner. This permits the interpretation of the limits in models predicting a narrow dielectron or dimuon resonance. A scan of different intrinsic width hypotheses is performed. Limits are set on the masses of various hypothetical particles. For the \ZPSSM (\ZPPSI) particle, which arises in the sequential standard model (superstring-inspired model), a lower mass limit of $\limitZssm\, (\limitZpsi)$\TeV is set at 95\% confidence level. The lightest Kaluza--Klein graviton arising in the Randall--Sundrum model of extra dimensions, with coupling parameters $k/\overline{M}_\mathrm{Pl}$  of 0.01, 0.05, and 0.10, is excluded at 95\% confidence level below \limitGone , \limitGtwo ,~and \limitGthree \TeV, respectively. In a simplified model of dark matter production via a vector or axial vector mediator, limits at 95\% confidence level are obtained on the masses of the dark matter particle and its mediator.
}

\hypersetup{%
pdfauthor={CMS Collaboration},%
pdftitle={Search for high-mass resonances in dilepton final state in proton-proton collisions at sqrt(s)=13 TeV},%
pdfsubject={CMS},%
pdfkeywords={CMS, physics, dileptons, resonance}}

\maketitle

\section{Introduction}
\label{sec:introduction}

Neutral resonances decaying to lepton pairs occur in a variety of theoretical models that attempt to extend the standard model (SM) of particle physics.
Grand unified theories (GUTs), including superstring and left-right-symmetric models (LR), achieve unification of the three forces at a high energy scale and predict the existence of new neutral gauge bosons~\cite{Leike:1998wr,Zp_SSM_1}.
These bosons might be light enough to be produced at current or future colliders.
Theories that allow the gravitational force to propagate into extra spatial dimensions~\cite{Randall:1999ee} could explain the large separation between the electroweak symmetry breaking energy scale and the gravitational energy scale.
In such models, graviton excitations could be observed as spin-2 high-mass resonances.
Moreover, high mass neutral resonances are predicted in models where dark matter (DM), whose existence is suggested by many astrophysical and cosmological observations~\cite{Ade:2015xua}, has a particle explanation.
In these theories, interactions between the DM and SM particles could be mediated by high-mass, weakly coupled particles.
Indirect evidence for DM could potentially be seen at the CERN LHC by searching for a high-mass DM mediator decaying to a dilepton final state.

Typical models predicting extra {\PZ}-like bosons extend the gauge group of the SM by additional $U^\prime(1)$ gauge groups.
The $U^\prime(1)$ gauge groups, or a linear combination of them, can be broken near the TeV scale, giving rise to new massive gauge bosons denoted as $\zp$.
A generalized version of these models uses a continuously varying angle to describe the mixing of the $U^\prime(1)$ generators.
The cross section for charged lepton pair production via a $\zp$ vector boson can, in the narrow-width approximation (NWA), be expressed in terms of the quantity $c_\PQu w_\PQu + c_\PQd w_\PQd$~\cite{Carena:2004xs,Accomando:2010fz}.
The parameter $c_\PQu$ ($c_\PQd$) contains information about the model-dependent $\zp$ boson couplings to the up-type (down-type) quarks, while $w_\PQu$ ($w_\PQd$) depends on the up-type (down-type) quark parton distribution functions (PDFs).
The parameterization of the linear mixing of the relevant $U^\prime(1)$ generators produces a contour in the ($c_\PQd,c_\PQu$) plane that represents each class of models.
Commonly considered models are the generalized sequential model (GSM)~\cite{Accomando:2010fz}, containing the $\ZPSSM$ boson that has SM-like couplings to SM fermions~\cite{Altar:1989}; GUT models based on the $E_6$ gauge group, containing the $\ZPPSI$ boson~\cite{Leike:1998wr,Zp_PSI_3}; and high-mass neutral bosons of the left (L)-right (R) symmetric extensions of the SM based on the $SU(2)_\text{L} \otimes SU(2)_\text{R} \otimes U(1)_\text{B-L}$ gauge group, where B-L refers to the difference between baryon and lepton numbers.
Specific choices of the mixing angle produce different models such as those shown in Table~\ref{tab:BenchmarkModels}, with their exact definition given in Ref.~\cite{Accomando:2010fz}.
The ($c_\PQd,c_\PQu$) plane parameterization provides a model-independent way to create a direct correspondence between the experimental bounds on $\zp$ production cross sections and the parameters of the Lagrangian.
The translation of the experimental limits into the ($c_\PQd,c_\PQu$) plane is studied both in the context of the NWA and by taking finite widths into account.
The two procedures, NWA and finite widths, have been shown to give the same results~\cite{Accomando:2010fz}.
A further study including the effects of interference~\cite{Accomando:2013sfa} has demonstrated that the two procedures can still be used with an appropriate choice of the invariant mass window, within which the cross section is calculated.

\begin{table*}[htb!]
\centering
\topcaption{
\label{tab:BenchmarkModels}
Various benchmark models with their corresponding mixing angles, their branching fraction ($\mathcal{B}$) to dileptons, the $c_{\PQu}$ and $c_{\PQd}$ parameter values and their ratio, and the width to mass ratio of the associated $\zp$ boson.}
\begin{tabular}{ccccccc}
$U^\prime(1)$ model & Mixing angle  & $\mathcal{B}(\ell^{+}\ell^{-}$)     & $c_{\PQu}$               &  $c_{\PQd}$              &  $c_{\PQu}$/$c_{\PQd}$     &  $\Gamma_\text{\zp}$/M$_\text{\zp}$ \\ \hline
E$_\text{6}$        &               &                  	 &			&			&						&	\\
U(1)$_{\chi}$       &     0         &     0.061            & $6.46\times10^{-4}$ & $3.23\times10^{-3}$ &  0.20                & 0.0117 			\\
U(1)$_{\psi}$       &     $0.5\pi$  &     0.044	         & $7.90\times10^{-4}$ & $7.90\times10^{-4}$ &  1.00                & 0.0053 			\\
U(1)$_{\eta}$       &   $-0.29\pi$  &     0.037            & $1.05\times10^{-3}$ & $6.59\times10^{-4}$ &  1.59                & 0.0064 		\\
U(1)$_\text{S}$     &   $0.129\pi$  &     0.066            & $1.18\times10^{-4}$ & $3.79\times10^{-3}$ &  0.31                & 0.0117 			\\
U(1)$_\text{N}$     &    $0.42\pi$  &     0.056            & $5.94\times10^{-4}$ & $1.48\times10^{-3}$ &  0.40                & 0.0064 		\\[2ex]
LR         	        &               &               	     &			           &			         &			            & 			\\
U(1)$_\text{R}$     &     0         &     0.048            & $4.21\times10^{-3}$ & $4.21\times10^{-3}$ &  1.00                & 0.0247 			\\
U(1)$_\text{B-L}$   &     0.5$\pi$  &     0.154            & $3.02\times10^{-3}$ & $3.02\times10^{-3}$ &  1.00                & 0.0150 			\\
U(1)$_\text{LR}$    &  $-0.128\pi$  &     0.025            & $1.39\times10^{-3}$ & $2.44\times10^{-3}$ &  0.57                & 0.0207 			\\
U(1)$_\text{Y}$     &    0.25$\pi$  &     0.125            & $1.04\times10^{-2}$ & $3.07\times10^{-3}$ &  3.39                & 0.0235 			\\[2ex]
GSM                 &               &                      &		               &			         &		                &	 \\
U(1)$_\text{SM}$    &  $-0.072\pi$  &     0.031            & $2.43\times10^{-3}$ & $3.13\times10^{-3}$ &  0.78                & 0.0297 			\\
U(1)$_\text{T3L}$   &    0          &     0.042            & $6.02\times10^{-3}$ & $6.02\times10^{-3}$ &  1.00                & 0.0450 			\\
U(1)$_\text{Q}$     &     0.5$\pi$  &     0.125            & $6.42\times10^{-2}$ & $1.60\times10^{-2}$ &  4.01                & 0.1225 			\\
\end{tabular}
\end{table*}

Searches for high-mass $\zp$ gauge bosons have been performed by the CMS Collaboration at the LHC with proton-proton collision data collected at $\sqrt{s} = 7\TeV$~\cite{Chatrchyan:2011wq,Chatrchyan:2012it} and with data collected at $8\TeV$~\cite{Chatrchyan:2012oaa,Khachatryan:2014fba}.
More recently CMS performed a search using the combination of 2015 data collected at $13\TeV$ with data collected at $8\TeV$~\cite{Khachatryan:2016zqb}.
Searches for high-mass $\zp$ gauge bosons have also been performed by the ATLAS Collaboration with data collected at $7\TeV$~\cite{Aad:2011xp,Aad:2012hf}, with data collected at $8\TeV$~\cite{Aad:2014cka}, and with data collected at $13\TeV$~\cite{Aaboud:2017buh}.

Kaluza--Klein graviton ($\GKK$) excitations arising in the Randall--Sundrum (RS) model of extra spatial dimensions~\cite{Randall:1999vf, Randall:1999ee} involve a finite five-dimensional bulk that is warped as a function of the position of the four-dimensional subspace in the fifth dimension.
In particular, the RS model predicts excited Kaluza--Klein modes of the graviton, without suppressing its couplings to the SM particles.
The modes appear as spin-2 resonances and can decay into dilepton final states.
There are two free parameters in the model: the mass of the first graviton excitation and the coupling $k/\overline{M}_\mathrm{Pl}$, where $k$ is the warp factor of the five-dimensional anti-de Sitter space and $\overline{M}_\mathrm{Pl}$ is the reduced Planck mass.
The intrinsic widths of the first excitation of the gravitons for the coupling parameters $k/\overline{M}_\mathrm{Pl}$ of 0.01, 0.05, and 0.10, are 0.01, 0.36 and 1.42\GeV, respectively.
Results of searches for resonances in $\Pp\Pp$ collision data have previously been reported by the ATLAS and CMS Collaborations~\cite{Chatrchyan:2012oaa,Aad:2014cka}.
At the Tevatron, the CDF and D0 Collaborations have published results based on a $\Pp\Pap$ collision sample at $\sqrt{s} = 1.96\TeV$, corresponding to an integrated luminosity of approximately 5\fbinv~\cite{CDF_Zp,CDF_RS,D0_RS,D0_Zp,CDF_SSM,CDF_RSele}.

We also consider a simplified model with a single DM particle that has sizeable interactions with the SM fermions through an additional spin-1 high-mass particle mediating the SM-DM interaction.
In this model, the mediator is a vector or an axial-vector boson and is exchanged in the $s$ channel~\cite{Albert:2017onk,Backovic:2015soa}.
There are five free parameters in this model: the DM mass $m_\text{DM}$, the mediator mass $m_\text{Med}$, the coupling $g_\text{DM}$ between the mediator and the DM particle, and the universal couplings $g_\ell$ and $g_{\PQq}$ between the mediator and the SM charged leptons and quarks, respectively.
These five parameters define the production rate of the mediator, its DM and leptonic/hadronic decay rates, and the kinematic distributions of the signal events.
We investigate two sets of benchmark coupling values that illustrate the complementary strengths of dijet and dilepton searches and the typical impact of searches for dilepton resonances in this model~\cite{Albert:2017onk}:
\begin{itemize}
\item   vector mediator with small couplings to leptons: $g_{\PQq}=0.1$, $g_\text{DM}=1.0$, $g_\ell=0.01$;
\item   axial-vector mediator with equal couplings to quark and leptons:  $g_\text{DM}=1.0$,\\ $g_{\PQq}=g_\ell=0.1$.
\end{itemize}
Possible interference between the mediator of the dilepton process and the Drell--Yan (DY) background is well below 5\% and can be safely neglected in the present analysis~\cite{Albert:2017onk}.

The results presented in this paper are obtained from an analysis of the data sample collected in 2016 at $\sqrt{s}=13\TeV$, corresponding to an integrated luminosity of \anaLumiee\fbinv for the dielectron channel and \anaLumimumu\fbinv for the dimuon channel.
The invariant mass spectra of the observed dilepton final states are scrutinized for possible deviations from the SM background predictions.
Background yields are estimated with simulated samples and normalized to their relative cross sections using the highest order calculations available.
The sum of these backgrounds is normalized to the observed yield in the dilepton invariant mass region of 60--120\GeV.
An exception to this is the background from multijet events, which is estimated from the data using control regions.
Additionally, the \ttbar background prediction from simulation is also cross-checked via \Pe$\mu$ events in data, as discussed in Sect.~\ref{sec:backgrounds}.

Limits are set on the ratio of the cross section for dilepton production via a new boson to the cross section for dilepton production via the SM \PZ boson.
This is done in order to remove the dependence on the CMS integrated luminosity measurement and to suppress the correlated uncertainties between the low- and high-mass regions.
The computation of the observed limit and significance involves an arbitrary choice of the intrinsic width of the new high-mass resonance.
The choice of the intrinsic width can potentially affect the statistical interpretation of the result, therefore we provide limits by scanning different width hypotheses, as discussed in Sect.~\ref{sec:results}.
The analysis is designed to minimize the effect of the specific model assumptions on the results, allowing the results to be interpreted in the framework of any high-mass resonance decaying into lepton pairs with a width and  pseudorapidity distribution similar to the reference model used.
The couplings of the $\zp$ model will not only impact the width of the peak region of the high-mass resonance but also the tails due to PDFs and interference effects between the SM electroweak bosons and the new resonance in DY $\cPZ{/}\gamma^* \to \Pep\Pem / \Pgmp\Pgmm$ processes.
In the NWA chosen for this analysis, the contributions to the signal cross section from PDFs and interference off-shell effects, which are highly model dependent, are removed.
Therefore, in this work we consider only the resonant peak, taking into account the effects of the intrinsic width of the high-mass resonance on the experimentally observed mass peak region.

This paper is structured as follows. Section~\ref{sec:detector} contains a short description of the CMS detector.
In Section~\ref{sec:samples}, we discuss the data sample and the Monte Carlo (MC) event generators used for the signal and background simulation.
The lepton reconstruction and event selection relevant for the $\zp$ boson search are discussed in Section~\ref{sec:selection}.
The background estimation methods are described in Section~\ref{sec:backgrounds}.
In Section~\ref{sec:results} the statistical method used to extract the results and the statistical treatment of the systematic uncertainties are explained.
Results are summarized in Section~\ref{sec:summary}.

\section{The CMS detector}
\label{sec:detector}

The central feature of the CMS detector is a superconducting solenoid providing an axial magnetic field of 3.8\unit{T} and enclosing an inner tracker, an electromagnetic calorimeter (ECAL), and a hadron calorimeter (HCAL).
The inner tracker is composed of a silicon pixel detector and a silicon strip tracker, and measures charged particle trajectories in the pseudorapidity range $\abs{\eta}<2.5$.
The ECAL and HCAL, each composed of a barrel and two endcap sections, extend over the range $\abs{\eta} < 3.0$.
The finely segmented ECAL consists of nearly 76\,000 lead tungstate crystals, while the HCAL is constructed from alternating layers of brass and scintillator.
Forward hadron calorimeters encompass $3.0<\abs{\eta}<5.0$.
The muon detection system covers $\abs{\eta}<2.4$ with up to four layers of gas-ionization detectors installed outside the solenoid and sandwiched between the layers of the steel flux-return yoke.
Additional detectors and upgrades of electronics were installed before the beginning of the 13\TeV data collection period in 2015, yielding improved reconstruction performance for muons relative to the 8\TeV data collection period in 2012.
The efficiency to reconstruct and select muons that result from \PZ boson decays and pass specific identification selection criteria, has increased by approximately $2\%$ between Run 1 and Run 2 as a result of these upgrades~\cite{1748-0221-12-01-C01048}.
A more detailed description of the CMS detector, together with a definition of the coordinate system used and the relevant kinematic variables, can be found in Ref.~\cite{Chatrchyan:2008zzk}.

The CMS experiment has a two-level trigger system.
The level-1 (L1) trigger~\cite{Khachatryan:2016bia}, composed of custom hardware processors, selects events of interest using information from the calorimeters and muon detectors and reduces the readout rate from the 40\unit{MHz} bunch crossing frequency to a maximum of 100\unit{kHz}.
The software based high-level trigger (HLT)~\cite{Khachatryan:2016bia} uses the full event information, including that from the inner tracker, to reduce the event rate to around the 1\unit{kHz} that is retained for further processing.

\section{Simulated data samples}
\label{sec:samples}

The dominant background in this search is the DY process.
The simulated DY background is generated with \POWHEG~v2~\cite{Nason:2004rx,Frixione:2007vw,Alioli:2010xd,Alioli:2008gx,Frixione:2007nw,Re:2010bp} from next-to-leading order (NLO) matrix elements using the NNPDF3.0~\cite{Ball:2014uwa} PDF set, and with \PYTHIA~8.205~\cite{Sjostrand:2014zea} for parton showering and hadronization.
For all simulated SM samples, the default tune for \PYTHIA, CUETP8M1~\cite{Khachatryan:2015pea}, is used.
The DY cross section at NLO is corrected to next-to-next-to-leading order (NNLO) in perturbative quantum chromodynamics (QCD) by using a dilepton invariant mass dependent $K$-factor according to the predictions of the \FEWZ~3.1.b2 program~\cite{Li:2012wna}.
In addition, these predictions incorporate missing EW corrections at NLO.
For the \FEWZ calculations, the LUXqed\_\-plus\_\-PDF4LHC15\_\-nnlo\_\-100~\cite{Manohar:2016nzj} PDF set is used in which the QCD PDFs based on the PDF4LHC~\cite{Butterworth:2015oua} set are combined with the photon PDFs to account for pure quantum electrodynamics effects.
Another nonresonant background arises from a $\gamma\gamma$ initial state via $t$ and $u$ channel processes.
The photon-induced (PI) process produces two leptons in the final state~\cite{Bourilkov:2016qum,Bourilkov:2016oet}.
This contribution is included in the $K$-factor that corrects the DY NLO cross section.

The \ttbar, $\cPqt\PW$ and $\PW\PW$ backgrounds are simulated using \POWHEG~v2, with parton showering and hadronization described by \PYTHIA~8.205.
The NNPDF3.0 PDF set is used for all these samples.
The \ttbar cross section is calculated at NNLO with \textsc{top++}~\cite{Czakon:2011xx} assuming a top quark mass of 172.5\GeV.
The inclusive diboson processes $\PW\PZ$, and $\PZ\PZ$ are simulated at leading order (LO) using the \PYTHIA~8.205 program along with the NNPDF3.0 PDFs.
The production of DY $\Pgt^{+}\Pgt^{-}$ and $\PW$+jets is simulated at LO with the \MGvATNLO version 2.2.2~\cite{Alwall:2014hca} program.
The PDFs are evaluated using the LHAPDF library~\cite{Whalley:2005nh,Bourilkov:2006cj,Buckley:2014ana}.

We use a sample of events in which a $\ZPPSI$~boson is generated with a mass of 3000\GeV, and RS samples with the graviton generated at different mass values from 250 to 4000\GeV.
In all samples the high-mass resonances decay to electron and muon pairs.
Both signal samples are generated using the \PYTHIA~8.205 program with the NNPDF3.0 PDFs.
The $\ZPPSI$ sample is used to create simulated peaks in the dilepton mass plots of Fig.~\ref{fig:massSpectra}.
However, we use DY samples to model the $\zp$ at high masses, since the dilepton behaviour in this region is identical in the two cases.

The presence of additional $\Pp\Pp$ interactions in the same or adjacent bunch crossing observed in data (pileup) is incorporated in simulated events by including overlapping $\Pp\Pp$ interactions.
MC samples are corrected to reproduce the pileup distribution as measured in data (pileup reweighting), with an average number of pileup interactions per proton bunch crossing of approximately 22 for the 2016 data sample.
The detector response is simulated using the \GEANTfour~\cite{Agostinelli:2002hh} package.

\section{Lepton reconstruction and event selection}
\label{sec:selection}

The electron and muon reconstruction algorithms and event selection criteria used in this high-mass dilepton search are mostly unchanged from the previous analysis~\cite{Khachatryan:2016zqb}.
However, the muon selection criteria were modified in both the online and offline selection in order to increase the efficiency in the high mass region, above 1\TeV.

Energy deposits in the ECAL are combined into clusters under the assumption that each local maximum represents a single particle.
Any clusters consistent with originating from a single particle that may have undergone bremsstrahlung emission are grouped together.
If a track from the nominal interaction point is geometrically associated with a cluster, this track together with the cluster form an electron candidate.
The angular information of the electron candidate is taken from the track.
The energy of the electron uses only the ECAL deposits and is not combined with the track momentum.
Electron candidates are required to have a transverse momentum $\pt>35\GeV$ and satisfy $\abs{\eta_C}<1.44$ (ECAL barrel region) or $1.57<\abs{\eta_C}<2.50$ (ECAL endcap region), where $\eta_C$ is the pseudorapidity of the cluster of ECAL deposits comprising the electron with respect to the nominal centre of the CMS detector.
The transition region $1.44<\abs{\eta_C}<1.57$ is excluded as it leads to lower-quality reconstructed clusters, owing mainly to services and cables exiting between the barrel and endcap calorimeters.

The electron candidates are also required to pass a set of dedicated high energy electron selection criteria~\cite{CMS-ele-paper}.
This selection requires that the lateral spread of deposits in the ECAL be consistent with that of a single electron, that the track be matched to the ECAL deposits and be consistent with a particle originating from the nominal interaction point, and that the associated energy in the HCAL around the electron direction be less than 5\% of the reconstructed energy of the electron, once noise and pileup are taken into account.
The selection also requires that the electron be isolated in a cone of radius $\Delta R = \sqrt{\smash[b]{(\Delta\eta)^2+(\Delta\phi)^2}} = 0.3$ in both the calorimeter and tracker~\cite{Khachatryan:2016zqb}.
Only well-measured tracks that are consistent with originating from the same vertex as the electron are included in the isolation sum.

The efficiency of the trigger to select events with two electrons passing the analysis selection requirements is 98.5\%, when the barrel (endcap) electrons satisfy $\pt>36$ (38)$\GeV$.
This primary trigger is monitored by a suite of higher-threshold triggers which have progressively fewer selection requirements, culminating in a trigger that simply requires 800\GeV of $\pt$ in the ECAL.
These triggers are included to minimize the chance that unexpected reconstruction problems cause a lower than expected efficiency in the primary trigger.

In the selection of dielectron pairs, at least one of the two electrons must be in the ECAL barrel region in order to reduce the background from multijet events.
This also allows the endcap-endcap events to be used as a control sample for the QCD background estimate.
Dielectron pairs are not required to be oppositely charged as this leads to a significant efficiency loss at high invariant mass~\cite{Chatrchyan:2011wq}.
As the electron energy is solely obtained from the calorimeter, an incorrectly measured charge does not impact the measured mass.
If there are multiple possible dielectron pairs, the pair with the two highest $\pt$ electrons is selected.

The efficiency to trigger, reconstruct, and select an electron pair with invariant mass equal to 1\TeV within the detector acceptance is 69 (65)\% for barrel-barrel (barrel-endcap) events.
The trigger efficiency is measured in data and the total efficiency is estimated using simulated DY events and validated using data measurements at the \PZ boson mass peak.
Using \PZ bosons, it is possible to probe the efficiencies up to electron $\pt$ of 500\GeV within a few percent precision.
The uncertainty in the efficiency is $\pm3$ ($\pm5$)\% in the barrel (endcap) for electrons coming from a dielectron pair with a mass $m_{\ell\ell} > 120\GeV$.
The simulated efficiency reproduces the energy evolution of the observed efficiency in the measurable region from 40 to 500\GeV.

A candidate muon pair at the L1 trigger is required to have at least one muon reconstructed with segments in the muon detectors and transverse momentum \pt above 22\GeV.
These muons are then required to have $\pt>50\GeV$ and $\abs{\eta}<2.4$ at the HLT.
A prescaled HLT path is used to extract the observed yield in the invariant mass region of the \cPZ~boson peak ($60<m_{\ell\ell}<120$\GeV) in order to construct the normalization factor to that region.
This prescaled trigger has a \pt threshold of 27\GeV with the same L1 requirements as the main trigger.

The trigger efficiency for dimuon events, where both muons have $\pt >53\GeV$, is parameterized using simulated DY events.
It is measured to be around 99.5\% if both muons are in the barrel ($\abs{\eta}<1.2$), and 99.0\% for events with one muon in the endcap ($\abs{\eta}>1.2$).
These efficiencies are validated as a function of muon \pt and $\eta$, using data events that pass the full offline selection criteria.
This is done using muons present in high mass dilepton or high-\pt \cPZ~boson events (free from background contributions) and other data sets (selected by electron, missing transverse momentum, or jet requirements).
The measurements are found to be in agreement for muons with \pt up to 1.5\TeV and an uncertainty of $\pm0.3$ ($\pm0.7$)\% for barrel-barrel (barrel-endcap and endcap-endcap) dimuon events over the full mass range is assigned.

High-\pt ($>$200\GeV) muon offline reconstruction uses dedicated algorithms~\cite{MUO-10-004-PAS} to take into account the effects of radiative processes of high-energy muon interactions with the detector material.
Muon candidates are required to have $\pt> 53\GeV$, to be within the region of $\abs{\eta}<2.4$, and to pass dedicated high-momentum identification selection criteria~\cite{Khachatryan:2014fba}.
Muon candidates are also required to pass isolation requirements.
The summed \pt of tracks within a cone of radius $\Delta R = 0.3$ around the candidate direction should be less than 10\% of the \pt of the candidate, excluding from summation the muon candidate under consideration.

Oppositely charged muon candidates passing the selection are combined to form dimuon candidates.
A $\chi^2$ fit is performed to the common vertex between the two muons to ensure that they originate from the same vertex.
This fit is required to have reduced $\chi^2<20$.
The angle between the directions of the two muon candidates is required to be less than $\pi-0.02$ in order to suppress cosmic ray backgrounds.
If there are multiple possible dimuon pairs, the pair with the two highest $\pt$ muons is selected.

Standalone muon reconstruction and identification efficiencies as a function of the muon momentum, both in data and simulation, were probed using high quality isolated inner tracks extrapolated to the muon system.
In the barrel region MC efficiencies are validated up to 1.2\TeV, while in the endcaps we observed a small discrepancy with respect to data at the order of 10\% for muon momentum of 3\TeV.
The efficiency ratios between data and MC obtained as a function of the single muon momentum are then propagated as a function of the dimuon mass.
It  leads to $-1.5$ ($-6.5$)\% one-sided uncertainties on a 4\TeV mass for barrel-barrel (barrel-endcap and endcap-endcap) events.
This represents the dominant uncertainty in the signal acceptance times efficiency.

The efficiency to trigger, reconstruct, and select a muon pair with invariant mass equal to 1\TeV within the detector acceptance is $92.7^{+0.3}_{-0.5}\%$ ($92.5^{+0.7}_{-2.7}\%$) for barrel-barrel (barrel-endcap and endcap-endcap) events.
The efficiency for each of the two muons is correlated as is the case for electron pairs.

The experimental dilepton mass resolution is determined from simulation as a function of the generated dilepton mass.
The simulated mass resolution measured in simulations is smeared in order to be comparable to \cPZ~boson events selected in data.
This smearing is performed for electrons in the entire \pt range while for the muon channel this smearing is performed as a function of the leading muon \pt up to 800 (450)\GeV in the barrel-barrel (barrel-endcap and endcap-endcap) events.
The experimental mass resolution is 1.0 (1.5)\% for barrel-barrel (barrel-endcap) electron pairs with a mass of 1\TeV.
No uncertainty is assigned to the electron pair mass resolution as we use the resolution measured in the low-mass region, which also represents the most pessimistic value for higher masses.
The resolution for muon pairs with a mass of 1\TeV is $3.0\pm0.5$ ($4.0\pm0.6$)\% for barrel-barrel (barrel-endcap and endcap-endcap) events.
We assign a 15\% systematic uncertainty in these values, based on measurements with different functions to parameterize the resolution for muon pairs.

The response of the detector to leptons might depend on the increasing dilepton invariant mass.
For electrons this could reflect a nonlinear response of the readout electronics.
There is no evidence for such effects in the current data.
The energy scale uncertainty at high \pt (400\GeV) is validated at the 2 (1)\% level for electrons in the barrel (endcaps).
Since the calorimeter is expected to behave linearly with energy we therefore assume this is true for a 2\TeV mass.
As the muon \pt increases, it becomes increasingly sensitive to the detector alignment.
New methods have been developed for the 2016 data to determine a potential bias from this source.
The usage of muon alignment position uncertainties has been added in the HLT and offline muon reconstruction algorithms in order to compensate for misalignment.
The curvature distributions of positive and negative muons ($q$/\pt) in data are compared to those obtained in simulation for different $\eta$ and $\phi$ ranges.
The mass scale is within 1 (3)\% for barrel-barrel (barrel-endcap and endcap-endcap) dimuon events for the entire mass range up to 2.3\TeV, covering the region where dimuon events are observed.

\section{Backgrounds}
\label{sec:backgrounds}

The dominant SM background arises from the DY process.
In addition to the DY process, \EE and \MM pairs can be produced in the PI process $\gamma\gamma \to \ell^+\ell^-$~\cite{Bourilkov:2016qum} from photons radiated by the incoming protons.
The PI process was studied through investigations of proton PDFs that include photon contributions~\cite{Bourilkov:2016qum,Bourilkov:2016oet}.
Although the relative contribution of PI processes increases with dilepton mass, the effect on the statistical analysis of the data was found to be negligible.
Uncertainties in the PDFs, in the contributions from PI processes, and in the NNLO corrections to the cross sections dominate the systematic uncertainty in the amount of estimated background.
This in turn dictates the uncertainty in the background shape.
The uncertainty due to the PDFs is assessed using the PDF4LHC prescription~\cite{Butterworth:2015oua} and is found to vary from 1.4 to 20\% as the dilepton mass increases from 0.2 to 6\TeV.

Other sources of background are real leptons from the top quark-antiquark (\ttbar), single top quark ($\cPqt\PW$), diboson ($\PW\PW$, $\PW\PZ$, and $\PZ\PZ$), and DY \TT processes.
The contribution of these sources is reduced at high invariant dilepton mass.
These backgrounds are estimated using simulated events.
For \ttbar, $\cPqt\PW$, diboson, and DY \TT production, the yield of $\Pe\Pgm$ final states produced should be approximately equal to the sum of $\Pe\Pe$ and $\mu\mu$ final states.
Therefore the simulation predictions in these channels are compared to data in the $\Pe\Pgm$ final state.
The predictions from data and simulation are in agreement within uncertainties and no further action is taken.

Multijet-enriched data control samples are used to evaluate the contribution of jets misidentified as electrons, as described in Ref.~\cite{Khachatryan:2014fba}.
Similarly, multijet data control samples are used to evaluate the probability for jets and nonisolated muons to be identified as isolated muons.
The contribution of misidentified jets to the total background is 1--3\%; therefore even with large uncertainties up to 50\%, it has a negligible effect on the statistical analysis of the data.

The contribution of cosmic ray background events is negligible in this analysis due to the event selection, and is $<$0.2\% for muons with $\pt>300\GeV$.

\begin{figure}[h]
\centering
\includegraphics[width=0.49\textwidth]{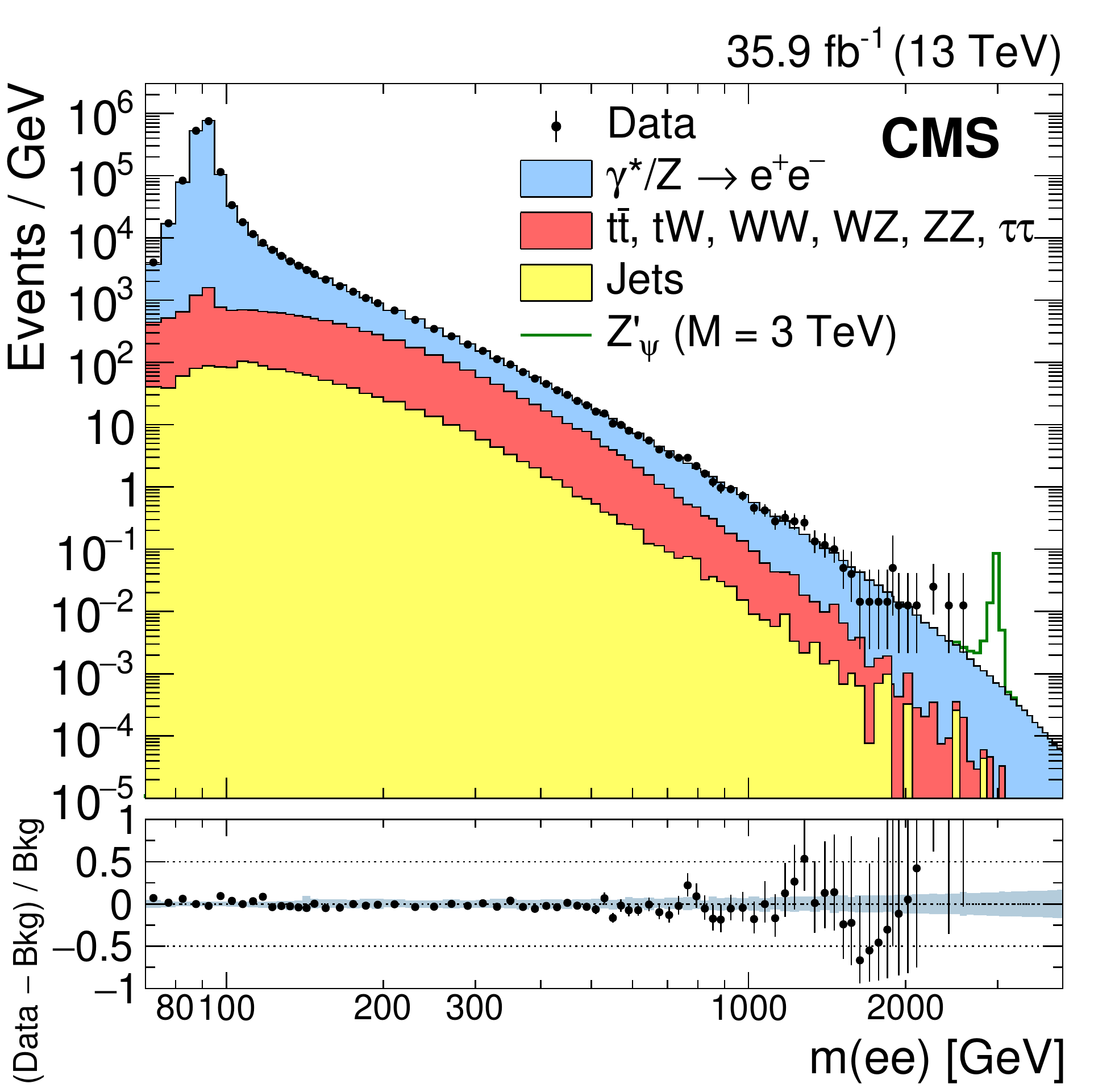}
\includegraphics[width=0.49\textwidth]{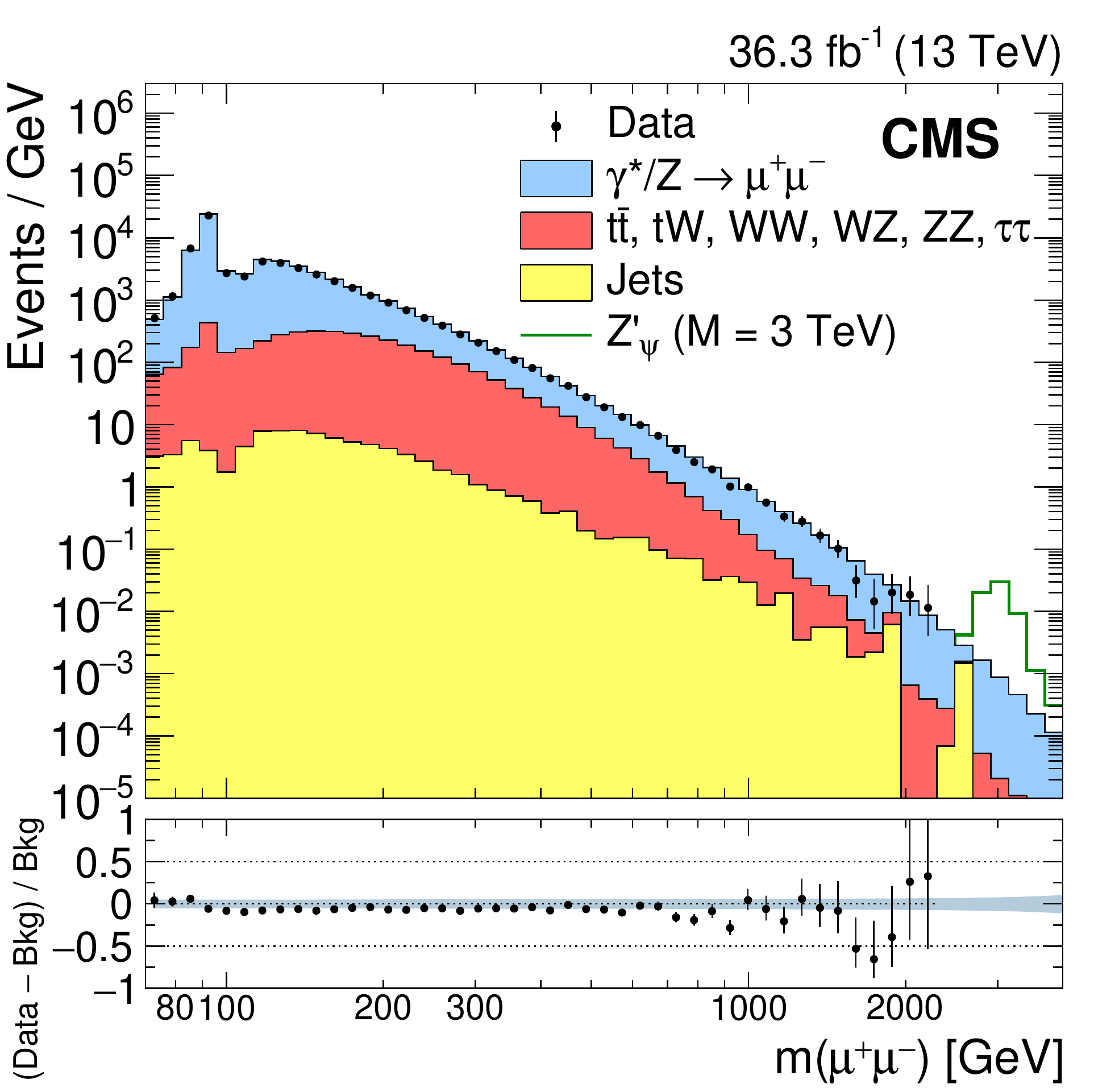}
\caption{
The invariant mass spectra of dielectron (left) and dimuon (right) events.
The points with error bars represent the observed yield.
The histograms represent the expectations from the SM processes.
The bins have equal width in logarithmic scale so that the width in GeV becomes larger with increasing mass.
Example signal shapes for a narrow resonance with a mass of 3\TeV are shown by the stacked open histograms.
The uncertainty bands in the ratio plots represent the systematic uncertainty in the background yields.
}
\label{fig:massSpectra}
\end{figure}

The observed invariant mass spectra of the dielectron and dimuon events are presented in Fig.~\ref{fig:massSpectra}.
The highest observed dilepton mass is 2.6\TeV and appears in the dielectron final state as a barrel-endcap event.
The highest observed dimuon mass is 2.3\TeV and appears as a barrel-endcap event.
The structure observed just above the \PZ boson peak in the dimuon channel is due to the high threshold ($\pt>53\GeV$) applied to the transverse momentum of the muon candidates.
The uncertainty bands in the ratio plots represent the systematic uncertainty in the background yields arising from the dilepton mass scale, trigger efficiency, acceptance times efficiency, \PZ boson normalization (1\% for dielectron and 5\% for dimuon channel), PDFs, non-DY background cross section determination (7\%), and lepton misidentification rates, summed in quadrature.
The selection criteria for the dielectron channel have a small sensitivity to pileup, so the contribution from pileup reweighting is included in the uncertainty band.
The dimuon selection criteria are by construction insensitive to pileup.

\begin{figure}[htbp]
\centering
\includegraphics[width=0.49\textwidth]{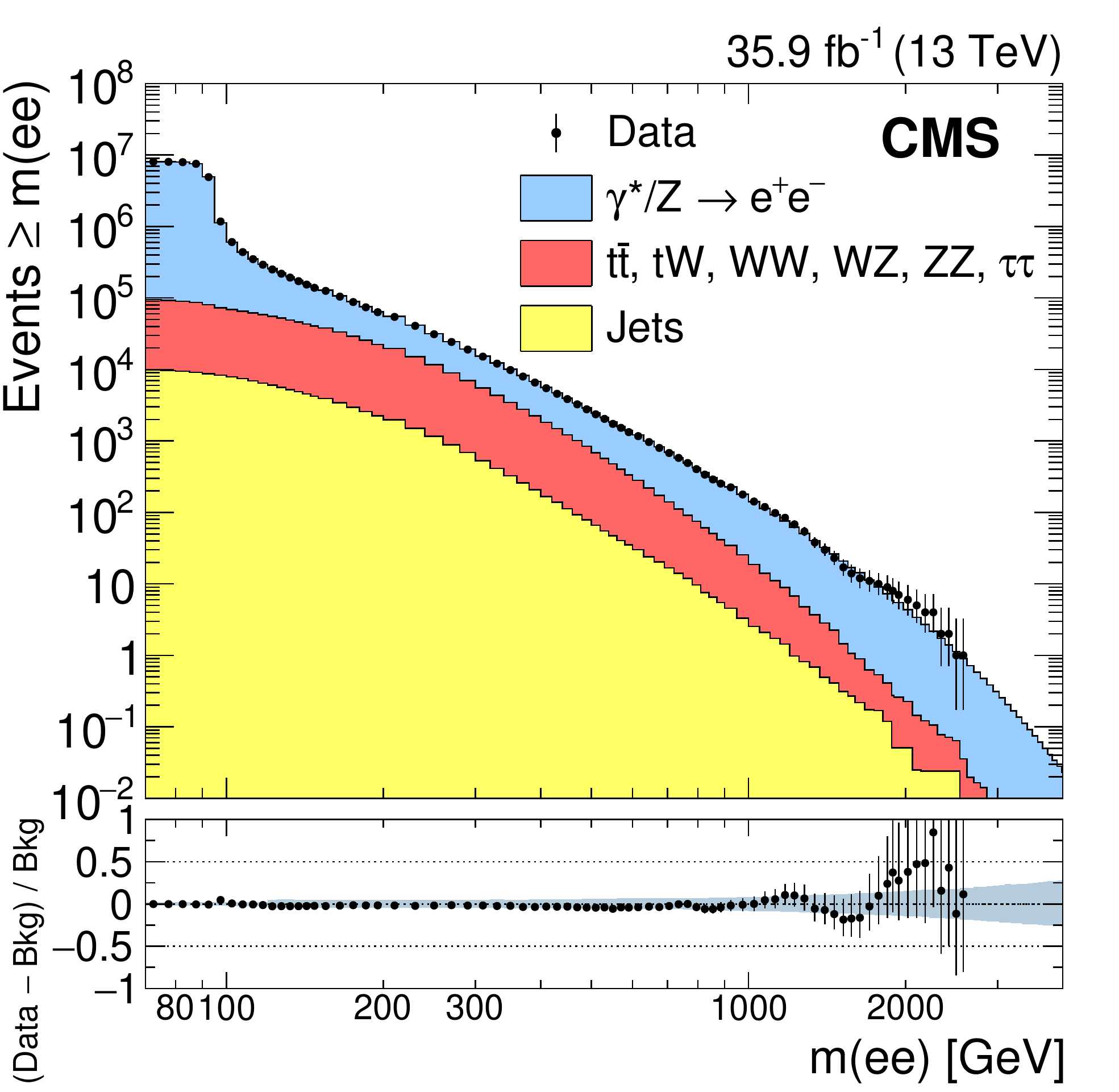}
\includegraphics[width=0.49\textwidth]{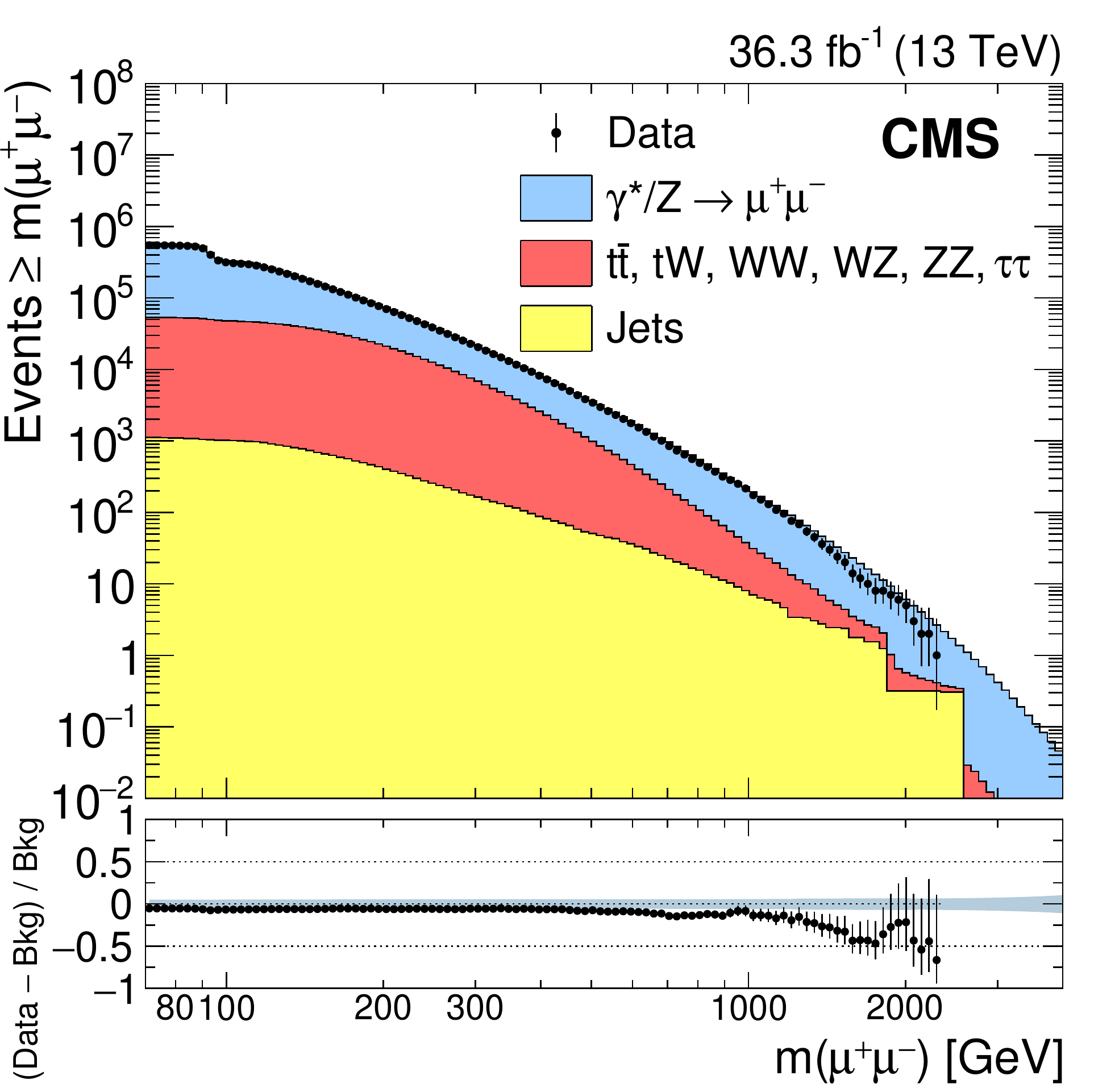}
\caption{
The cumulative distributions, where all events above the specified mass on the $x$ axis are summed, of the invariant mass spectra of dielectron (left) and dimuon (right) events.
The points with error bars represent the observed yield.
The histograms represent the expectations from SM processes.
The uncertainty bands in the ratio plots represent the systematic uncertainty in the background yields.
}
\label{fig:cum_spectra}
\end{figure}

The corresponding cumulative distributions are shown in Fig.~\ref{fig:cum_spectra}.
The SM expected yields in various mass bins are compared to the observed yields in Tables~\ref{tab:event_yieldee} and \ref{tab:event_yieldmumu}.
Observations agree with expectations in the entire mass region for the dielectron events.
A deficit of dimuon events is observed in the high-mass region compared to the expectations from SM processes (as shown on the right plot of Fig.~\ref{fig:cum_spectra}).
This deficit appears in the barrel ($\abs{\eta}<1.2$).
In the barrel-barrel category we observe two dimuon candidates in the region $m_{\mu\mu} > 1600$\GeV, where we expect ten dimuon candidates from MC simulations.
This leads to a local significance of the discrepancy equal to 2.9 standard deviations (s.d.).
This significance is reduced to 1.8 s.d. when considering the entire pseudorapidity range and is considered to be compatible with a statistical fluctuation.
No experimental sources for the deficit were identified.

\begin{table*}[htbp]
\centering
\topcaption{
The number of dielectron events in various invariant mass ranges.
The total background is the sum of the events for the SM processes listed.
The yields from simulation are normalized relative to the expected cross sections, and overall the simulation is normalized to the observed yield using the number of events in the mass window 60--120\GeV.
Uncertainties include both statistical and systematic components, summed in quadrature.
}
\begin{tabular}{ccr@{\hspace{1.5pt}}c@{\hspace{1.5pt}}l|r@{\hspace{1.5pt}}c@{\hspace{1.5pt}}lr@{\hspace{1.5pt}}c@{\hspace{1.5pt}}lr@{\hspace{1.5pt}}c@{\hspace{1.5pt}}l}
\multicolumn{1}{c}{$m_{\Pe\Pe}$ range} & Observed    & \multicolumn{3}{c}{Total}             & \multicolumn{3}{c}{$\cPZg$}           &	\multicolumn{3}{c}{$\ttbar$ + other}	&	\multicolumn{3}{c}{Jet mis-}	\\
\multicolumn{1}{c}{[{GeV}]}             & yield       & \multicolumn{3}{c}{background}        & \multicolumn{3}{c}{}			 		 &	\multicolumn{3}{c}{backgrounds}		&	\multicolumn{3}{c}{reconstruction}     \\ \hline
120--400   & $245\,101$     & $252\,000$  & $\pm$ & $13\,000$ & $199\,000$ & $\pm$ & $11\,000$      & $47\,700$ & $\pm$ & $2\,100$    & 5800 & $\pm$ & 2900    \\
400--600   & 4297           & 4430        & $\pm$ & 230       & 2890       & $\pm$ & 150            & 1400      & $\pm$ & 88          & 137  & $\pm$ & 69      \\
600--900   & 943            & 986         & $\pm$ & 64        & 739        & $\pm$ & 49             & 221       & $\pm$ & 17          & 26   & $\pm$ & 13      \\
900--1300  & 182            & 187         & $\pm$ & 14        & 156        & $\pm$ & 12             & 26.8      & $\pm$ & 2.3         & 3.9  & $\pm$ & 1.9      \\
1300--1800 & 33             & 34.3        & $\pm$ & 3.4       & 30.9       & $\pm$ & 3.2            & 2.8       & $\pm$ & 0.5         & 0.6  & $\pm$ & 0.3    \\
$>$1800    & 9              & 7.5         & $\pm$ & 1.1       & 7.0        & $\pm$ & 1.1            & 0.30      & $\pm$ & 0.04        & 0.13 & $\pm$ & 0.07    \\
\end{tabular}
\label{tab:event_yieldee}
\end{table*}

\begin{table*}[htb!]
\centering
\topcaption{
The number of dimuon events in various invariant mass ranges.
The total background is the sum of the events for the SM processes listed.
The yields from simulation are normalized relative to the expected cross sections, and overall the simulation is normalized to the observed yield using the number of events in the mass window 60--120\GeV , acquired using a prescaled low threshold trigger.
Uncertainties include both statistical and systematic components, summed in quadrature.
}
\begin{tabular}{c@{\hspace{6pt}}c@{\hspace{6pt}}r@{\hspace{1.5pt}}c@{\hspace{1.5pt}}l@{\hspace{6pt}}|r@{\hspace{1.5pt}}c@{\hspace{1.5pt}}l@{\hspace{10pt}}r@{\hspace{1.5pt}}c@{\hspace{1.5pt}}l@{\hspace{0pt}}r@{\hspace{1.5pt}}c@{\hspace{1.5pt}}l}
\multicolumn{1}{c}{$m_{\MM}$ range} & Observed    & \multicolumn{3}{c}{Total}             & \multicolumn{3}{c}{$\cPZg$}           &	\multicolumn{3}{c}{$\ttbar$ + other}	&	\multicolumn{3}{c}{Jet mis-}	   \\
\multicolumn{1}{c}{[{GeV}]}         & yield       & \multicolumn{3}{c}{background}        & \multicolumn{3}{c}{}			      &	\multicolumn{3}{c}{backgrounds}			&	\multicolumn{3}{c}{reconstruction} \\ \hline
120--400   & $244\,277$ & $260\,000$ & $\pm$ & $14\,000$ & $218\,000$ & $\pm$ & $11\,000$ & $40\,900$ & $\pm$ & $3\,500$ & \hspace{10pt} 800 & $\pm$ & 400   \\
400--600   &     5912   & 6290       & $\pm$ & 350       & 4340       & $\pm$ & 230       & 1900      & $\pm$ & 160      & 50  & $\pm$ & 25    \\
600--900   &     1311   & 1430       & $\pm$ & 80        & 1070       & $\pm$ & 60        & 340       & $\pm$ & 30       & 20  & $\pm$ & 10    \\
900--1300  &      244   & 268        & $\pm$ & 15        & 220        & $\pm$ & 12        & 41        & $\pm$ & 4        & 7   & $\pm$ & 4      \\
1300--1800 &       41   & 50         & $\pm$ & 3         & 42.6       & $\pm$ & 2.5       & 5.4       & $\pm$ & 0.9      & 2.1   & $\pm$ & 1.1   \\
$>$1800    &        8   & 12.1       & $\pm$ & 1.5       & 9.8        & $\pm$ & 0.7       & 1.1       & $\pm$ & 0.4      & 1.2 & $\pm$ & 0.6   \\
\end{tabular}
\label{tab:event_yieldmumu}
\end{table*}

\section{Statistical analysis, results, and interpretation}
\label{sec:results}

The mass distributions are scrutinized for possible deviations from the SM background predictions.
No significant deviations are observed.
The limits are expressed as a function of $R_{\sigma}$, which is the ratio of the cross section for dilepton production via a \PZpr boson to the measured cross section for dilepton production via the \PZ boson in the mass window 60--120\GeV:
\begin{equation}
\label{eq:rsigma}
R_\sigma = \frac{\sigma(\Pp\Pp\to \cPZpr+X\to\ell\ell+X)}
                {\sigma(\Pp\Pp\to \cPZ+X  \to\ell\ell+X)}.
\end{equation}
Expressing the limits as a ratio, reduces the dependency on the theoretical prediction of the \PZ\ boson cross section as well as the correlated experimental uncertainties.

For the electron and muon channel combination, the branching fractions of these two channels are assumed to be the same.
The signal cross section corresponds to that obtained in the narrow width approximation; specifically, off-shell contributions from PDFs and interference effects are not included.
The limits are set using a Bayesian method with an unbinned extended likelihood function~\cite{Khachatryan:2014fba} using the framework developed for statistically combining Higgs boson searches~\cite{CMS-NOTE-2011-005}, which is based on the \textsc{RooStats} package~\cite{Moneta:2010pm}.
The signal probability density function (pdf) used is a convolution of a Breit--Wigner (BW) function and a Gaussian function with exponential tails to either side (Cruijff~\cite{delAmoSanchez:2010ae}).
The BW function models the intrinsic width of the particle, while the Cruijff function models the detector response.
A Crystal Ball (CB) function~\cite{Oreglia:1980cs} better describes the signal shape at higher dielectron masses compared to the Cruijff function.
Therefore for $m_{\Pe\Pe}>2300\GeV$, a CB function is used.
The background pdf for the dielectron channel is provided by the following formula:
\begin{equation}
\label{eq:shapebckgELE}
\begin{split}
m^{\kappa}\exp\Bigl(\sum\limits_{i=0}^3 \alpha_{i} m^{i}\Bigr),\qquad \text{if $m \le 600\GeV$} \\
m^{\lambda}\exp\Bigl(\sum\limits_{i=0}^3 \beta_{i} m^{i}\Bigr),\qquad \text{if $m > 600\GeV$},
\end{split}
\end{equation}
while for the dimuon channel the following functional form is used:
\begin{equation}
\label{eq:shapebckgMUO}
\begin{split}
m^{\mu}\exp\Bigl(\sum\limits_{i=0}^2 \gamma_{i} m^{i}\Bigr),\qquad \text{if $m \le 500\GeV$} \\
m^{\nu}\exp\Bigl(\sum\limits_{i=0}^3 \delta_{i} m^{i}\Bigr),\qquad \text{if $m > 500\GeV$}.
\end{split}
\end{equation}
The different background pdfs for dielectrons and dimuons reflect the different background composition in the two channels.
For each final state, the parameters of the background pdf are obtained by fitting the total background distribution produced using SM MC generators and the background arising from misidentified jets deduced from the data.
The fits to the background distribution are set for masses above 120\GeV.

For the signal cross section we use a positive uniform prior.
The systematic uncertainties in the dilepton mass originating from efficiencies, resolution and scale, as discussed in Sect.~\ref{sec:selection}, are treated as nuisance parameters and assigned log normal priors.
The relative mass scale of the different channels is the only uncertainty with a noticeable impact.
The limits are calculated in a mass window of $\pm$6 times the signal width, with this window being symmetrically enlarged until there is a minimum of 100 data events in it.
This procedure sets the level of the statistical uncertainty in the local background amplitude; the level is chosen to dominate the expected systematic uncertainties in the background shape at high mass.
The total background uncertainty ranges from approximately 3\% at 200\GeV to 12\% at 5\TeV for both electrons and muons.
At low mass it is driven by normalization uncertainties; while at high mass, PDFs and higher-order corrections are the dominant sources of uncertainty.

\begin{figure}[htb]
\centering
\includegraphics[width=0.49\textwidth]{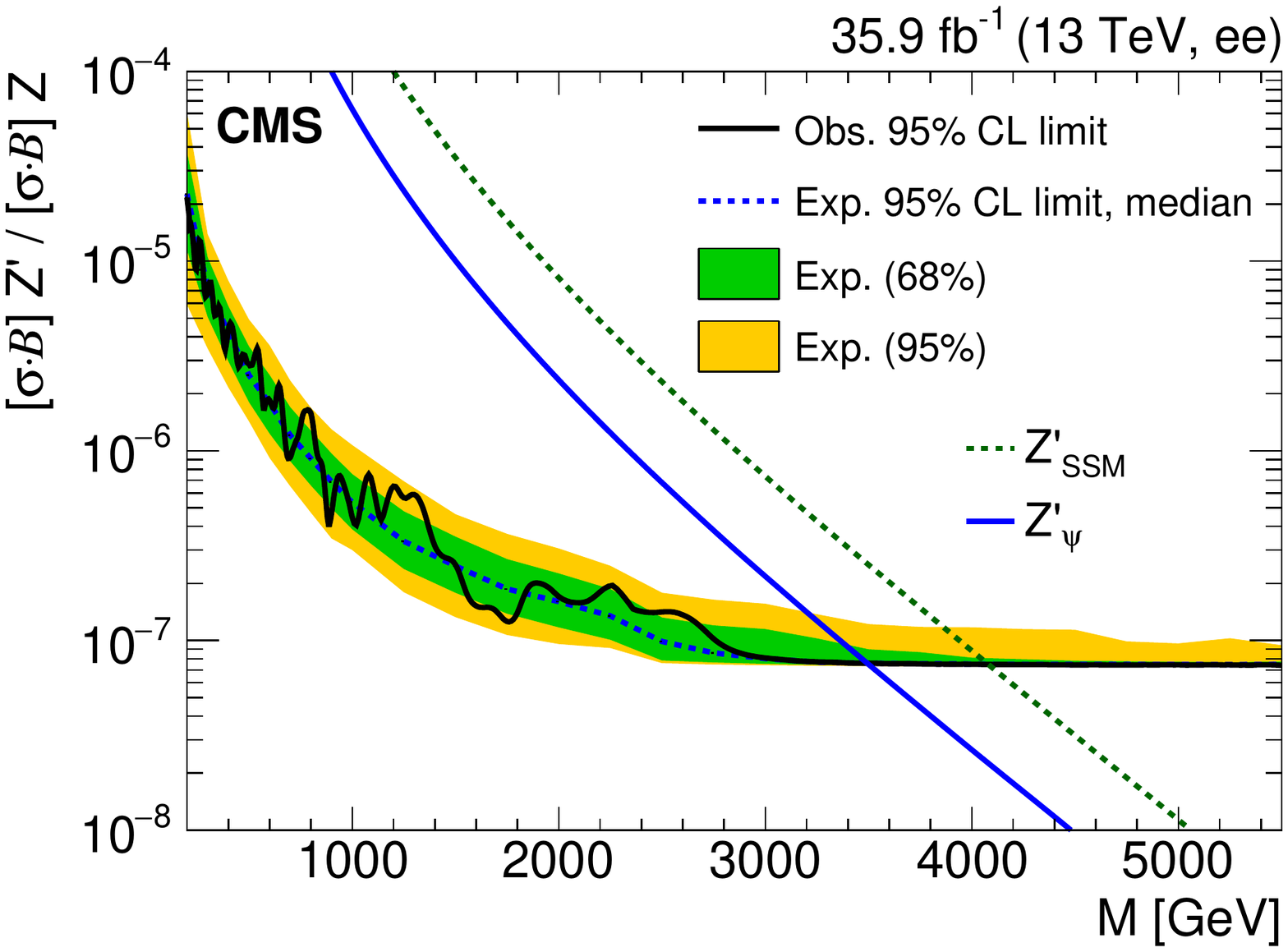}
\includegraphics[width=0.49\textwidth]{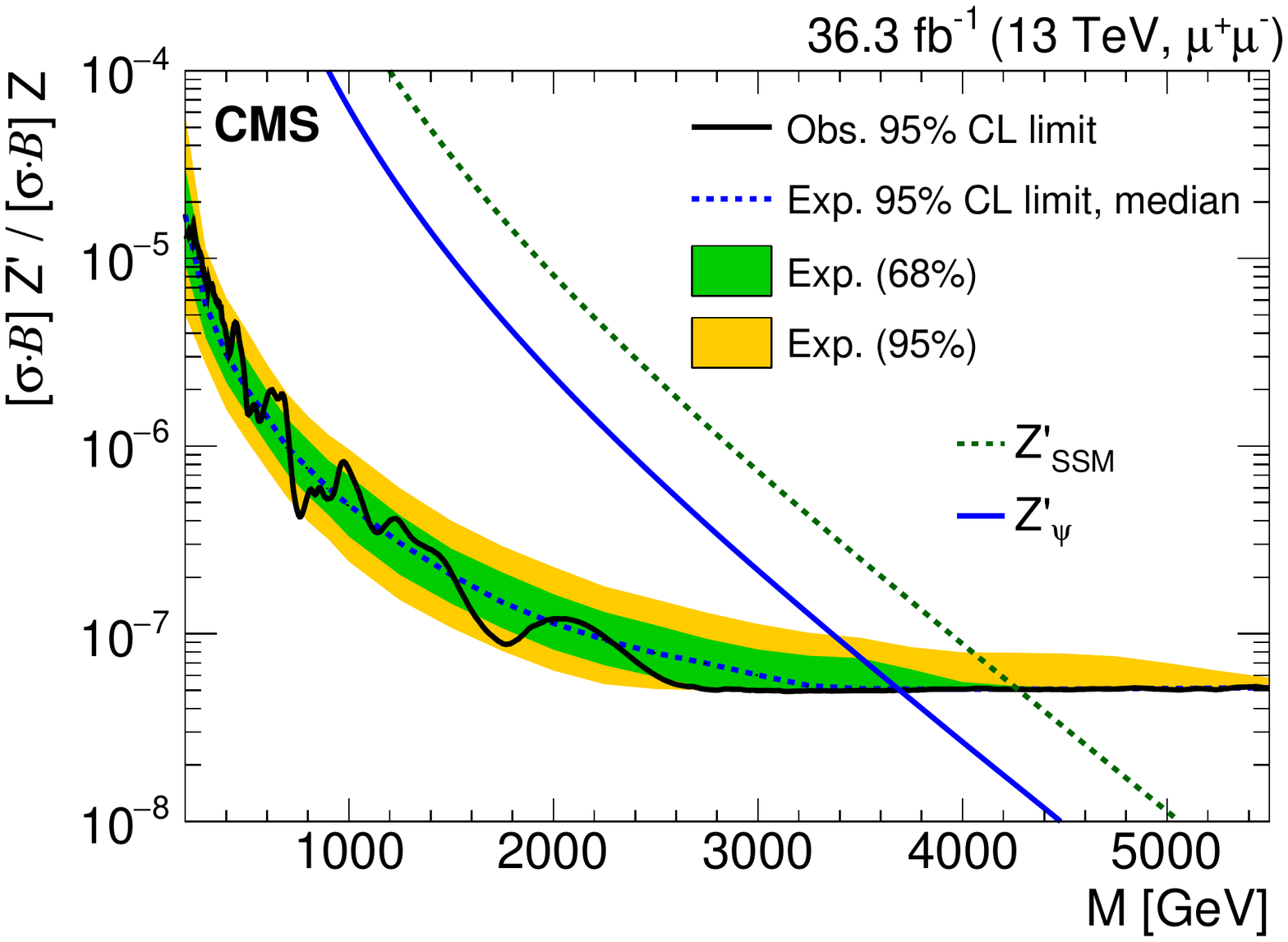}
\includegraphics[width=0.49\textwidth]{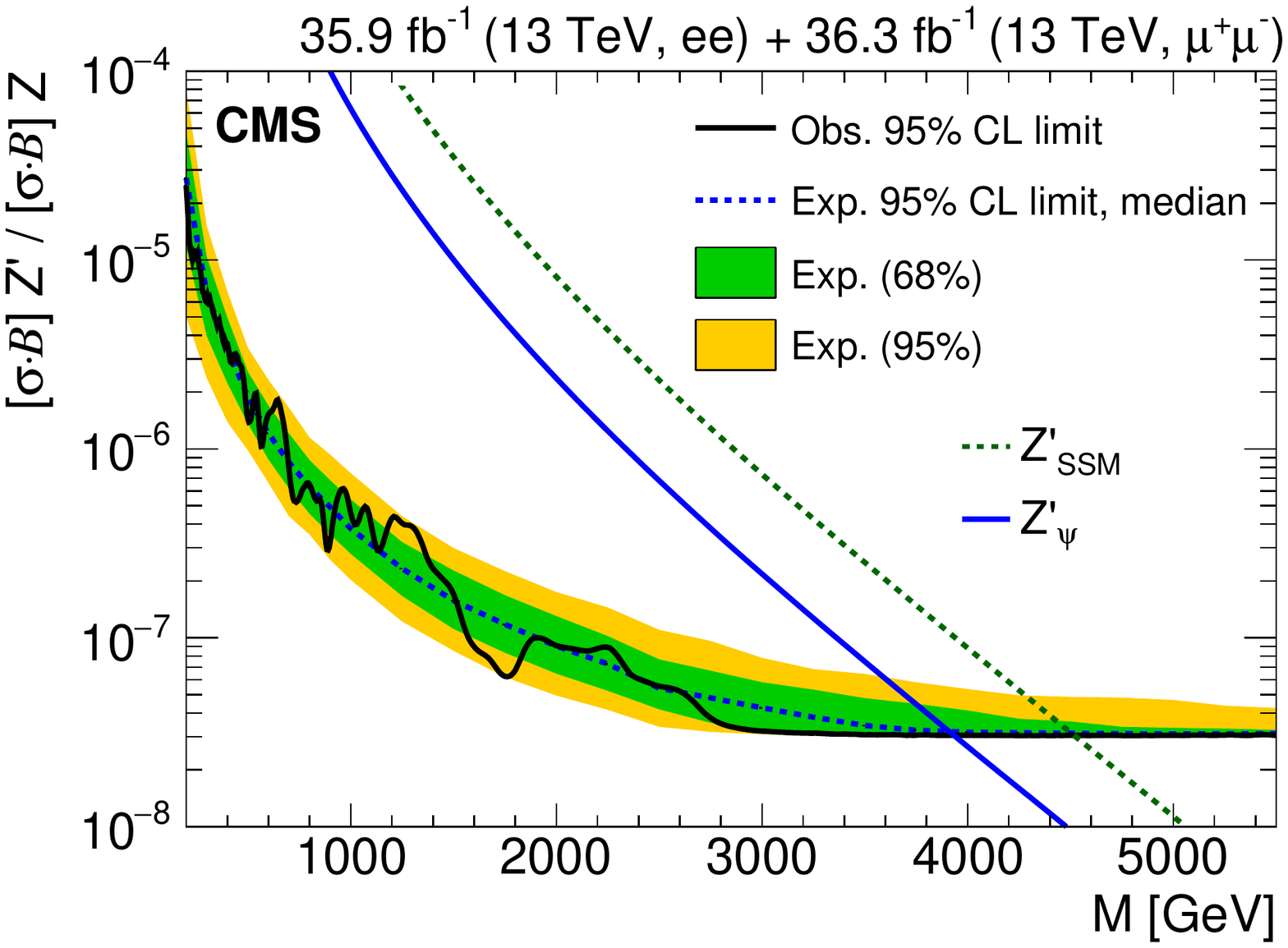}
\caption{
The upper limits at 95\% \CL on the product of production cross section and branching fraction for a spin-1 resonance with a width equal to 0.6\% of the resonance mass, relative to the product of production cross section and branching fraction of a \PZ boson, for the dielectron channel (left), dimuon channel (right), and their combination (lower).
The shaded bands correspond to the 68 and 95\% quantiles for the expected limits.
Theoretical predictions for the spin-1 $\ZPSSM$ and $\ZPPSI$ resonances are shown for comparison.
}
\label{fig:limits2}
\end{figure}

\begin{figure}[htb]
\centering
\includegraphics[width=0.49\textwidth]{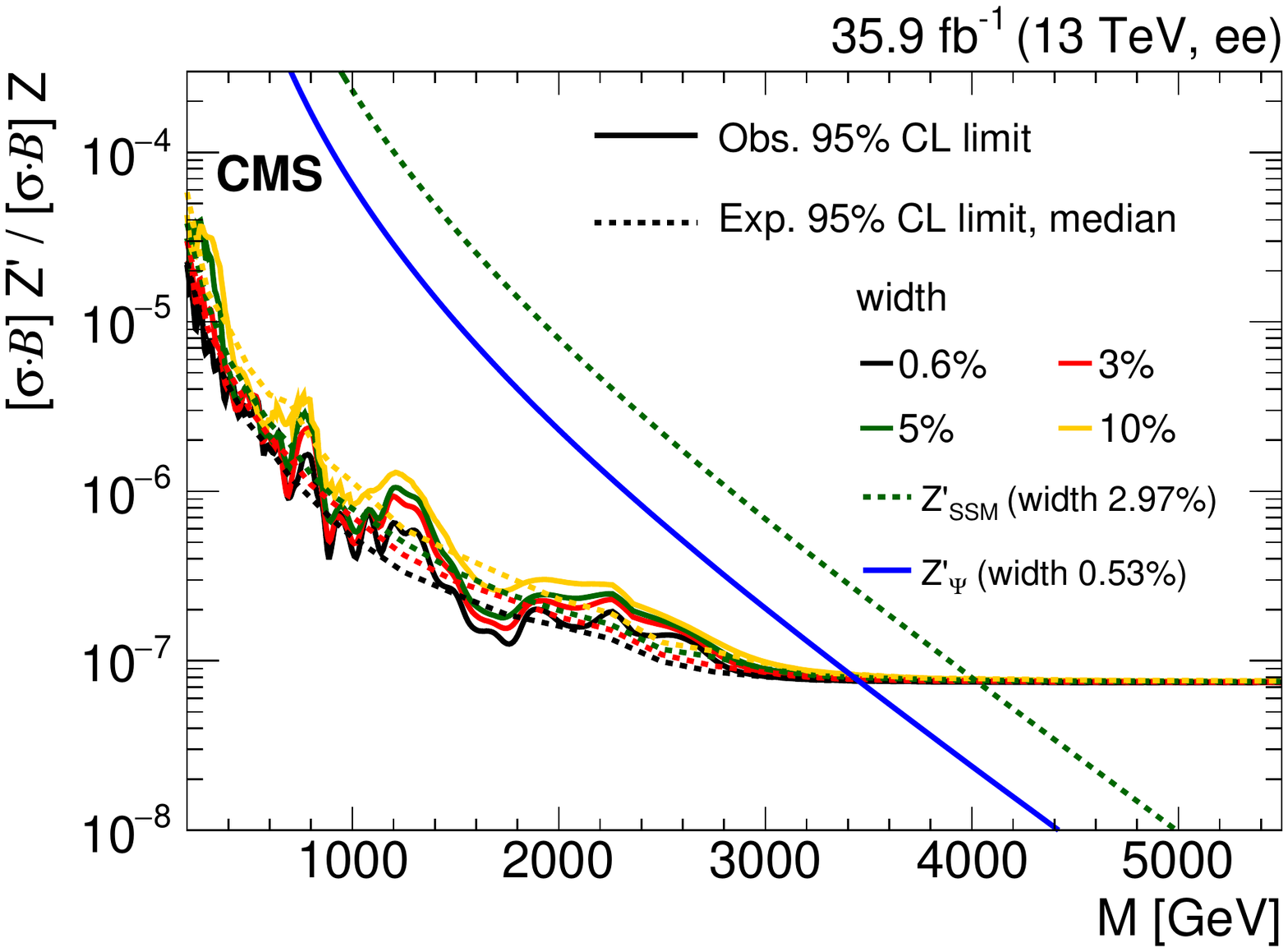}
\includegraphics[width=0.49\textwidth]{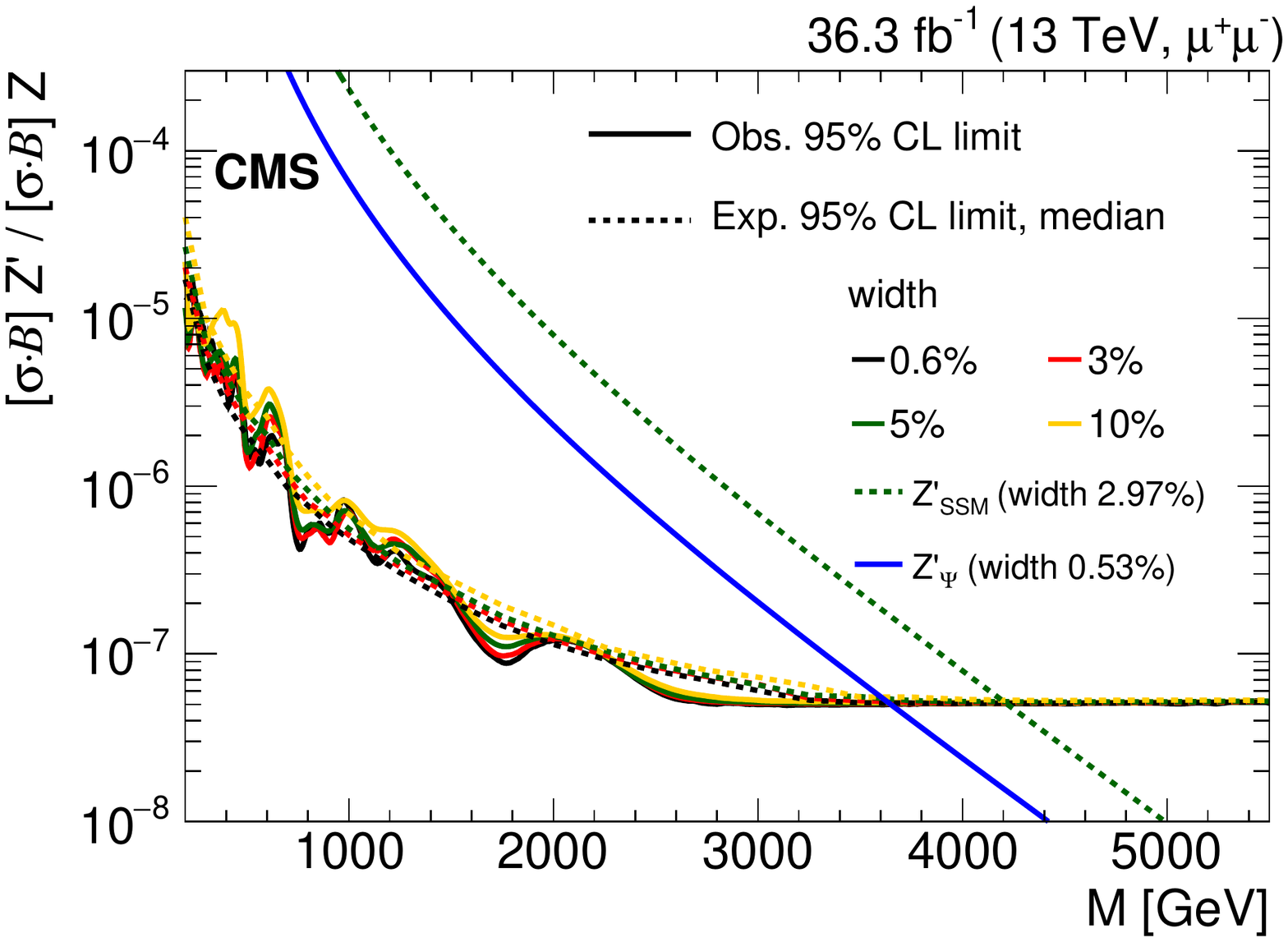}
\includegraphics[width=0.49\textwidth]{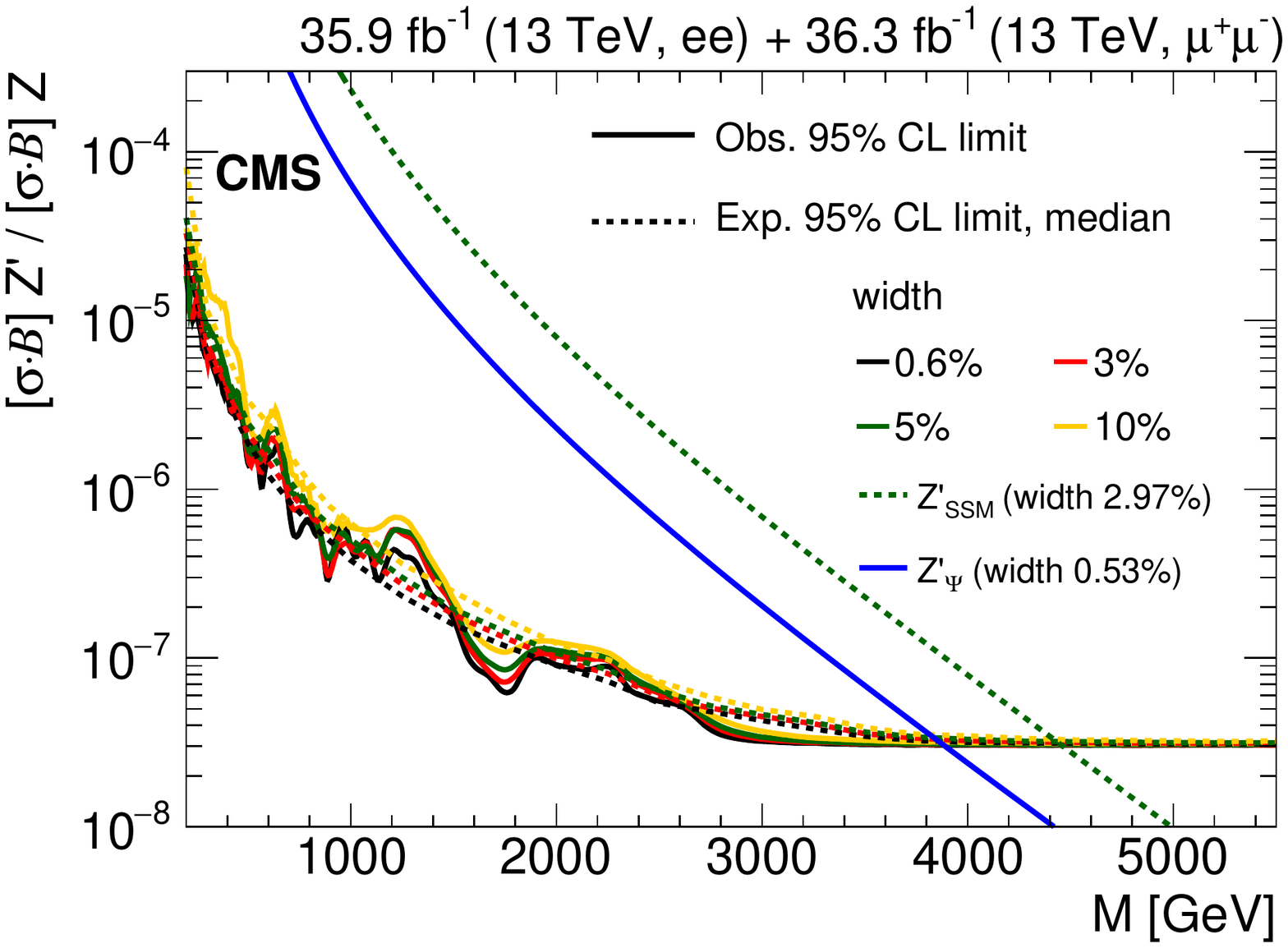}
\caption{
The upper limits at 95\% \CL on the product of production cross section and branching fraction for a spin-1 resonance, for widths equal to 0.6, 3, 5, and 10\% of the resonance mass, relative to the product of production cross section and branching fraction for a \PZ boson, for the dielectron channel (left), dimuon channel (right), and their combination (lower).
Theoretical predictions for the spin-1 $\ZPSSM$ and $\ZPPSI$ resonances are also shown.
}
\label{fig:limits3}
\end{figure}

The expected and observed limits for a resonance width equal to 0.6\% of the resonance mass are shown in Fig.~\ref{fig:limits2} for the dielectron channel, dimuon channels, and their combination.
Table~\ref{tab:massLimitsSpin1} presents the observed and expected 95\% confidence level (CL) lower limits on the masses of spin-1 $\ZPSSM$ and $\ZPPSI$ bosons.
Results for widths equal to 0.6, 3, 5 and 10\% of the resonance mass are shown in Fig.~\ref{fig:limits3} for the dielectron channel, dimuon channel, and their combination.
For masses below 2\TeV the expected limits become less stringent with increasing resonance widths.
At high masses, however, the experimental mass resolution dominates and the limits do not exhibit any dependence on the assumed resonance width.
Compared to the default width of 0.6\%, an increased width results in less stringent observed and expected limits, and a smoother variation of the observed limit as a function of mass.

The $R_{\sigma}$ curves are shown on the plots to obtain mass limits for \PZpr\ signal models.
The curves are constructed by dividing the LO cross section of a given model, calculated using the \PYTHIA~8.2 program with the NNPDF2.3~\cite{Ball:2012cx} PDFs, by the NNLO \PZ\ boson cross section of $1928\pm73\unit{pb}$ obtained with \FEWZ~3.1~\cite{Li:2012wna}.
As the limits presented here are set on the on-shell cross section and the \PYTHIA event generator includes off-shell effects, the cross section is calculated in a mass window of $\pm5\%$~$\sqrt{s}$ centred on the resonance mass, following the prescription of Ref.~\cite{Accomando:2013sfa}.
The validity of this procedure for the $\ZPSSM$ and $\ZPPSI$ bosons was explicitly checked in Ref.~\cite{Accomando:2013sfa} and is found to be accurate to approximately 5--7\%.
To account for NNLO QCD effects, the LO cross sections are multiplied by a mass independent $K$-factor.
The value of the $K$-factor is estimated at a dilepton mass of $4.5\TeV$ and found to be consistent with unity.
Applying a mass dependent $K$-factor, the $\ZPPSI$ resonance mass limit differs by only $50\GeV$, justifying the use of the simpler mass independent $K$-factor.

For the $\ZPSSM$ and $\ZPPSI$ bosons, we obtain 95\% \CL lower mass limits of $\limitZssm$ and $\limitZpsi\TeV$, respectively.
Recent measurements from the ATLAS experiment, based on 36.1\fbinv of proton-proton collision data collected at $\sqrt{s}=13\TeV$ in 2015 and 2016~\cite{Aaboud:2017buh}, have obtained 95\% \CL lower mass limits of 4.5 and 3.8$\TeV$ for the $\ZPSSM$ and $\ZPPSI$ bosons, respectively.

\begin{table}[!hbt]
\centering
\topcaption{
The observed and expected 95\% \CL lower limits on the masses of spin-1 $\ZPSSM$ and $\ZPPSI$ bosons, assuming a signal width of 0.6\% (3.0\%) of the resonance mass for $\ZPPSI$ ($\ZPSSM$).
}
\begin{tabular}{ccccc}
\multirow{2}{*}{Channel}  & \multicolumn{2}{c}{$\ZPSSM$} & \multicolumn{2}{c}{$\ZPPSI$}  \\
                          & Obs. [\TeVns{}] & Exp. [\TeVns{}]      & Obs. [\TeVns{}]  & Exp. [\TeVns{}]      \\\hline
\Pe\Pe                    &  4.10      & 4.10            & 3.45        & 3.45            \\
$\mu^+\mu^-$              &  4.25      & 4.25            & 3.70        & 3.70            \\
\Pe\Pe ~+ $\mu^+\mu^-$    &  4.50      & 4.50            & 3.90        & 3.90            \\
\end{tabular}
\label{tab:massLimitsSpin1}
\end{table}

\begin{figure}[htb]
\centering
\includegraphics[width=0.65\textwidth]{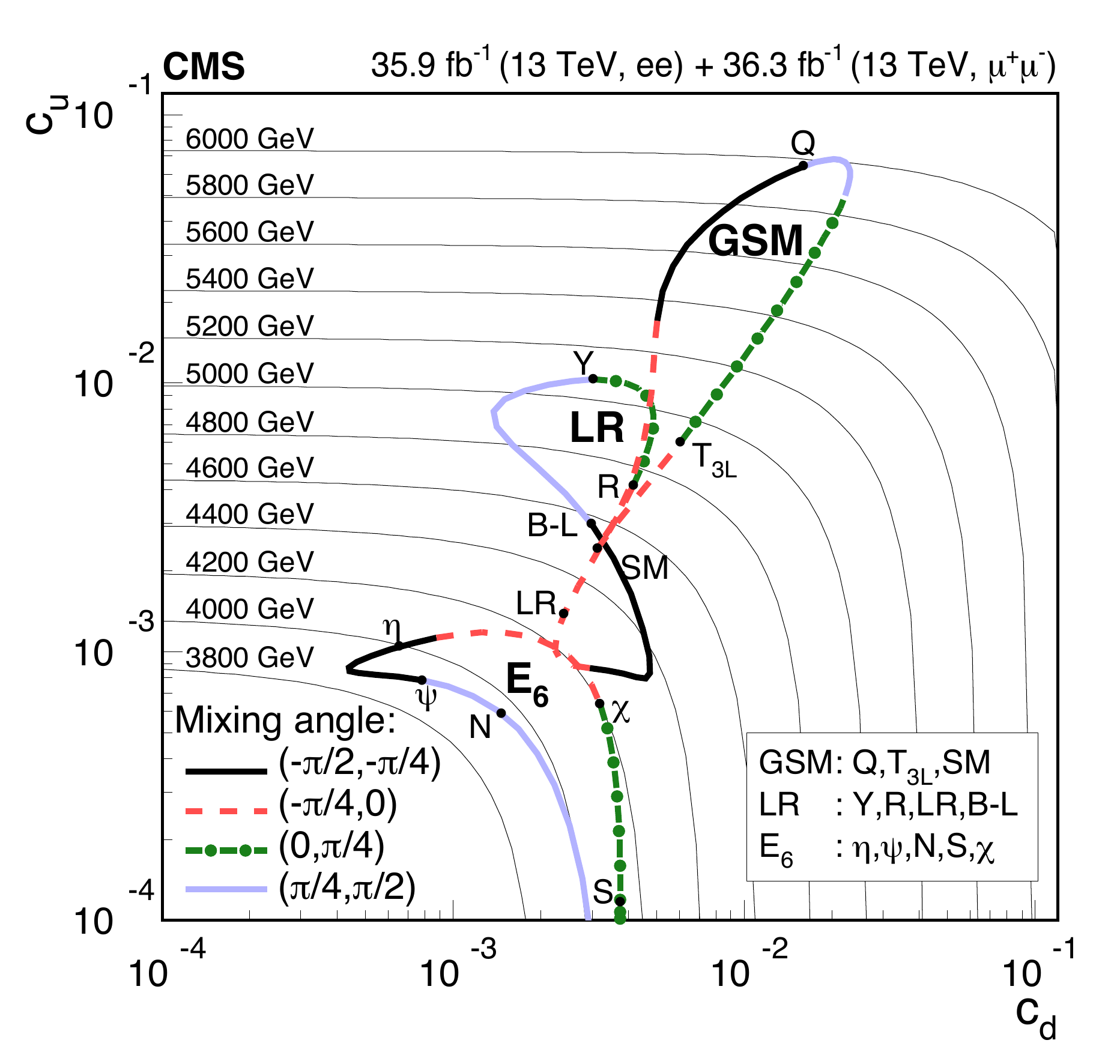}
\caption{
Limits in the $(c_{\PQd},c_{\PQu})$ plane obtained by recasting the combined limit at 95\% \CL on the $\zp$ boson cross section from dielectron and dimuon channels.
For a given $\zp$ boson mass, the cross section limit results in a solid thin black line.
These lines are labelled with the relevant $\zp$ boson masses.
The closed contours representing the GSM, LR, and E$_6$ model classes are composed of thick line segments.
Each point on a segment corresponds to a particular model, and the location of the point gives the mass limit on the relevant $\zp$ boson.
As indicated in the bottom left legend, the segment line styles correspond to ranges of the particular mixing angle for each considered model.
The bottom right legend indicates the constituents of each model class.
}
\label{fig:CuCd}
\end{figure}

In Fig.~\ref{fig:CuCd}, the cross section limit curve from Fig.~\ref{fig:limits2} is translated into the $(c_{\PQd},c_{\PQu})$ plane.
The LO cross section is a linear function of $c_{\PQd}$ and $c_{\PQu}$ ($\sigma_\text{LO} \propto c_{\PQd}w_{\PQd} + c_{\PQu}w_{\PQu}$, where $w_{\PQd}/w_{\PQu}$ is in the range 0.5--0.6 for the results shown here) and hence a single value of the cross section is represented by a straight line in the $(c_{\PQd},c_{\PQu})$ plane.
In the log-log plot shown in Fig.~\ref{fig:CuCd}, the thin lines labelled with a mass value correspond to the cross section limit at that mass.
The closed contours representing the GSM, LR, and E$_6$ model classes are composed of thick line segments which correspond to ranges of the particular mixing angle for each considered model.
A brief description of the models is given in Section~\ref{sec:introduction} with further information provided in Table~\ref{tab:BenchmarkModels} and the exact definition of the models discussed in Ref.~\cite{Accomando:2010fz}.
The mass limit on the relevant $\zp$ boson in any model, where $c_{\PQd}$ and $c_{\PQu}$ have been determined, can be read off this plot.

\begin{figure*}[htb]
\centering
\includegraphics[width=0.49\textwidth]{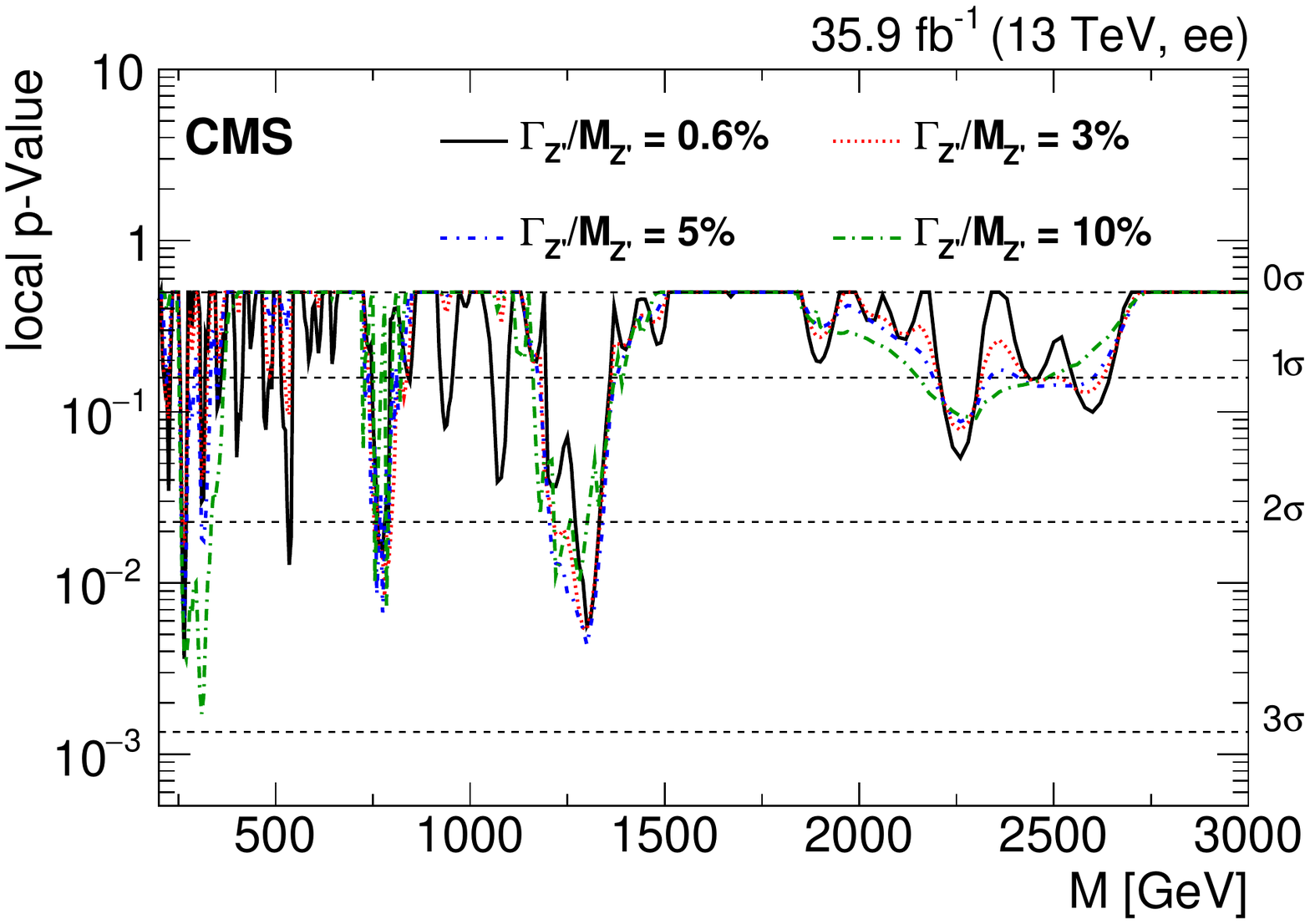} \hfill
\includegraphics[width=0.49\textwidth]{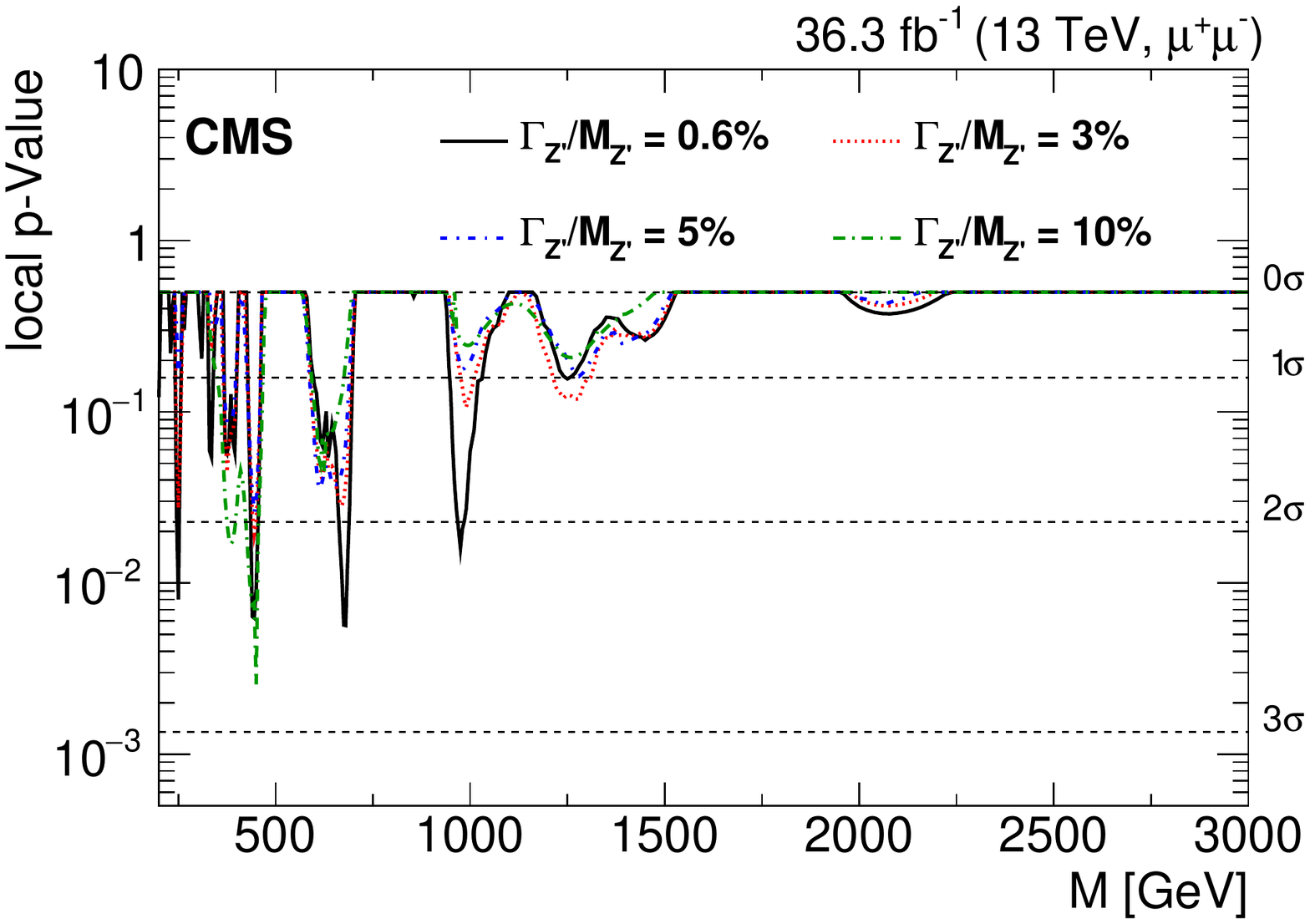}
\includegraphics[width=0.49\textwidth]{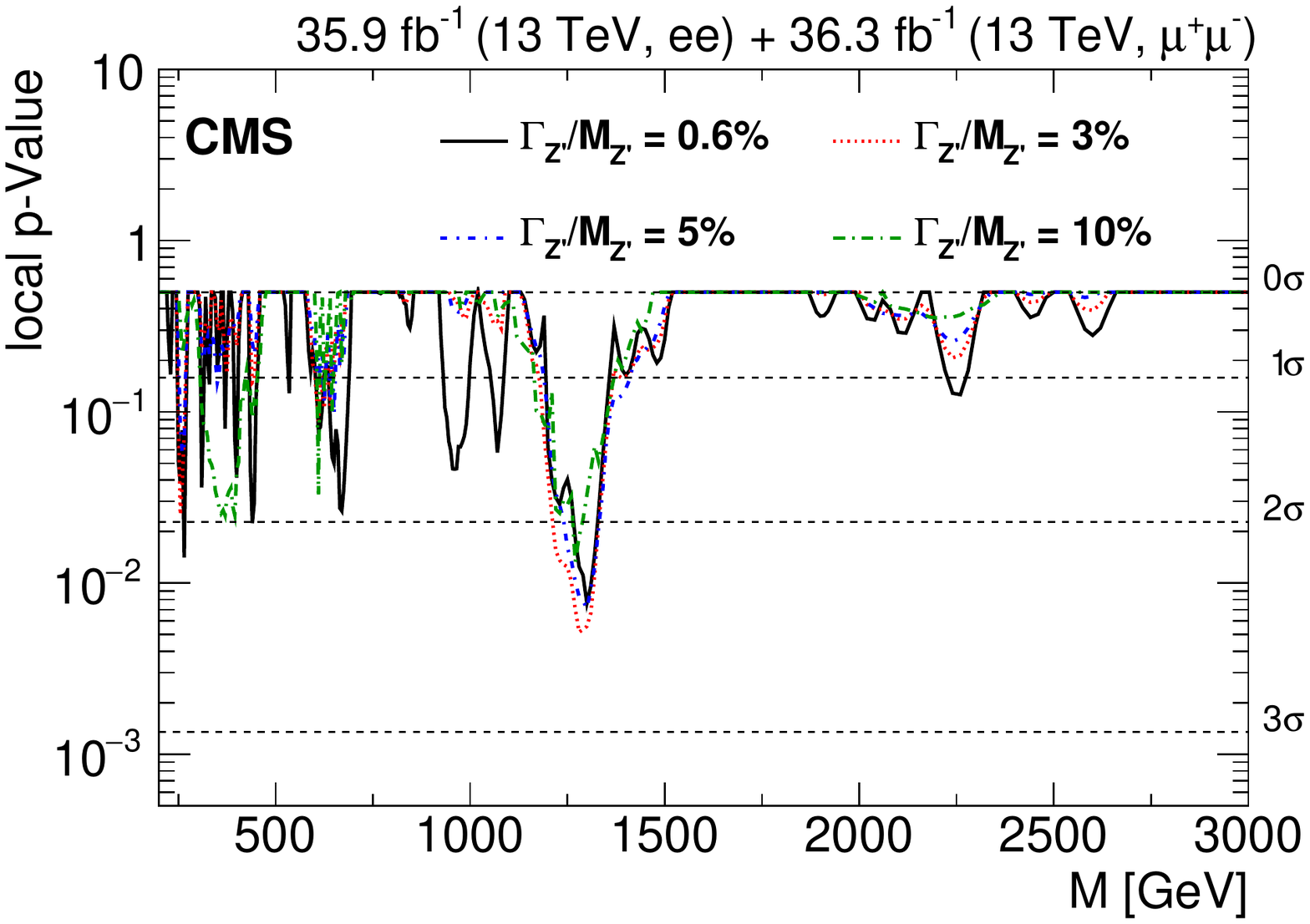}
\caption{
The observed local $p$-value for the dielectron channel (left), dimuon channel (right), and their combination (lower) as a function of the dilepton invariant mass.
}
\label{fig:pvalue}
\end{figure*}

For completeness we quantify a possible presence of an excess of events over what is expected for the background by computing the $p$-value.
The $p$-value for different signal width hypotheses is shown for both the separate and combined channels in Fig.~\ref{fig:pvalue}.
The largest excess in the combined result is observed around $M=1300\GeV$ having a local significance of around 2.5 s.d. for a spin-1 resonance with widths 0.6 to 5.0\%.
This corresponds to a global significance of $-0.92$ s.d. after taking into consideration the look elsewhere effect~\cite{Gross:2010qma} in the mass range 200 to 5500\GeV.
The global significance is expressed as the corresponding number of standard deviations using the one-sided Gaussian tail convention.
The methodology of the $p$-value computation is described in Ref.~\cite{Chatrchyan:2012xdj}.

\begin{figure}[htb]
\centering
\includegraphics[width=0.49\textwidth]{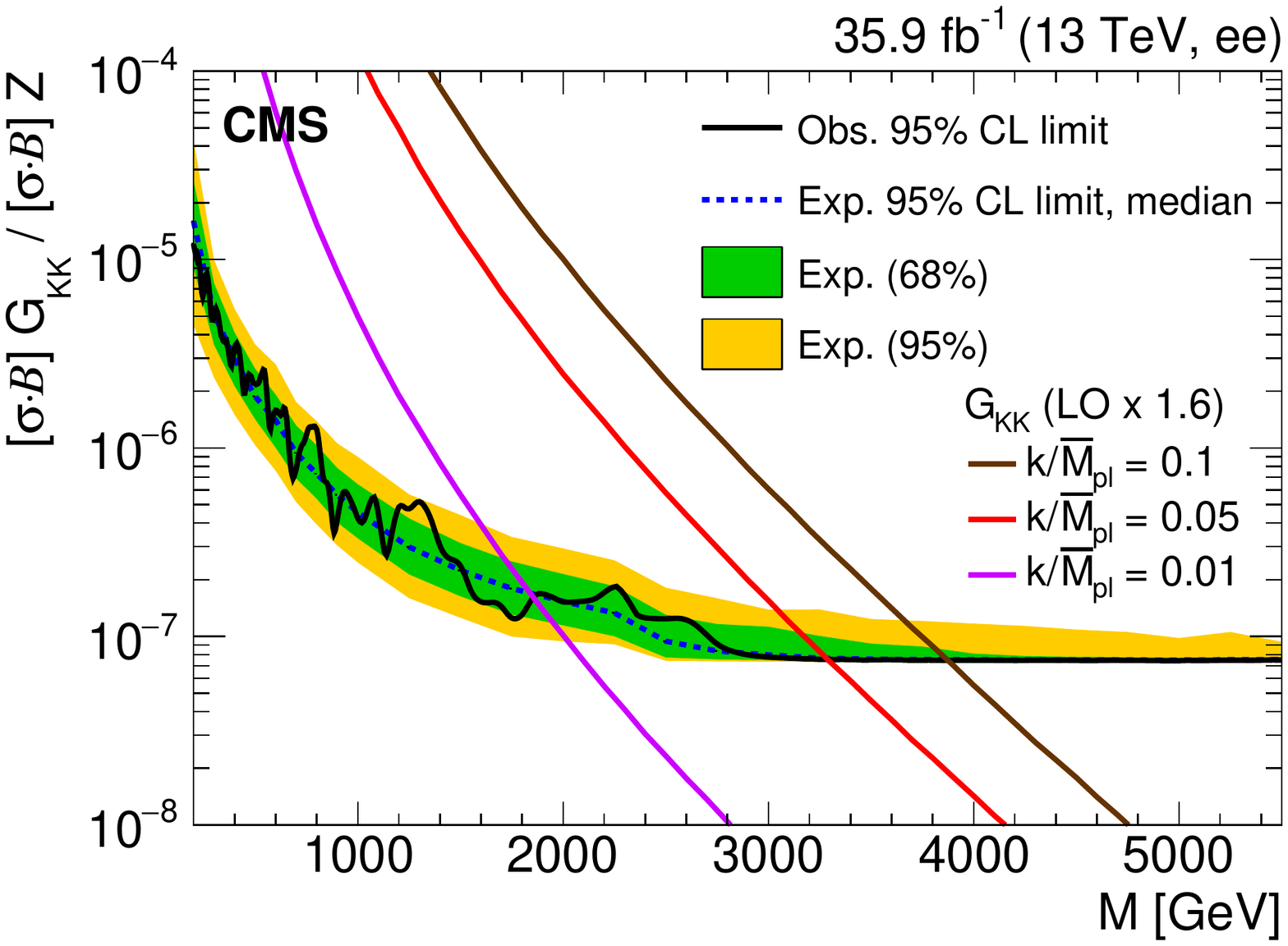}
\includegraphics[width=0.49\textwidth]{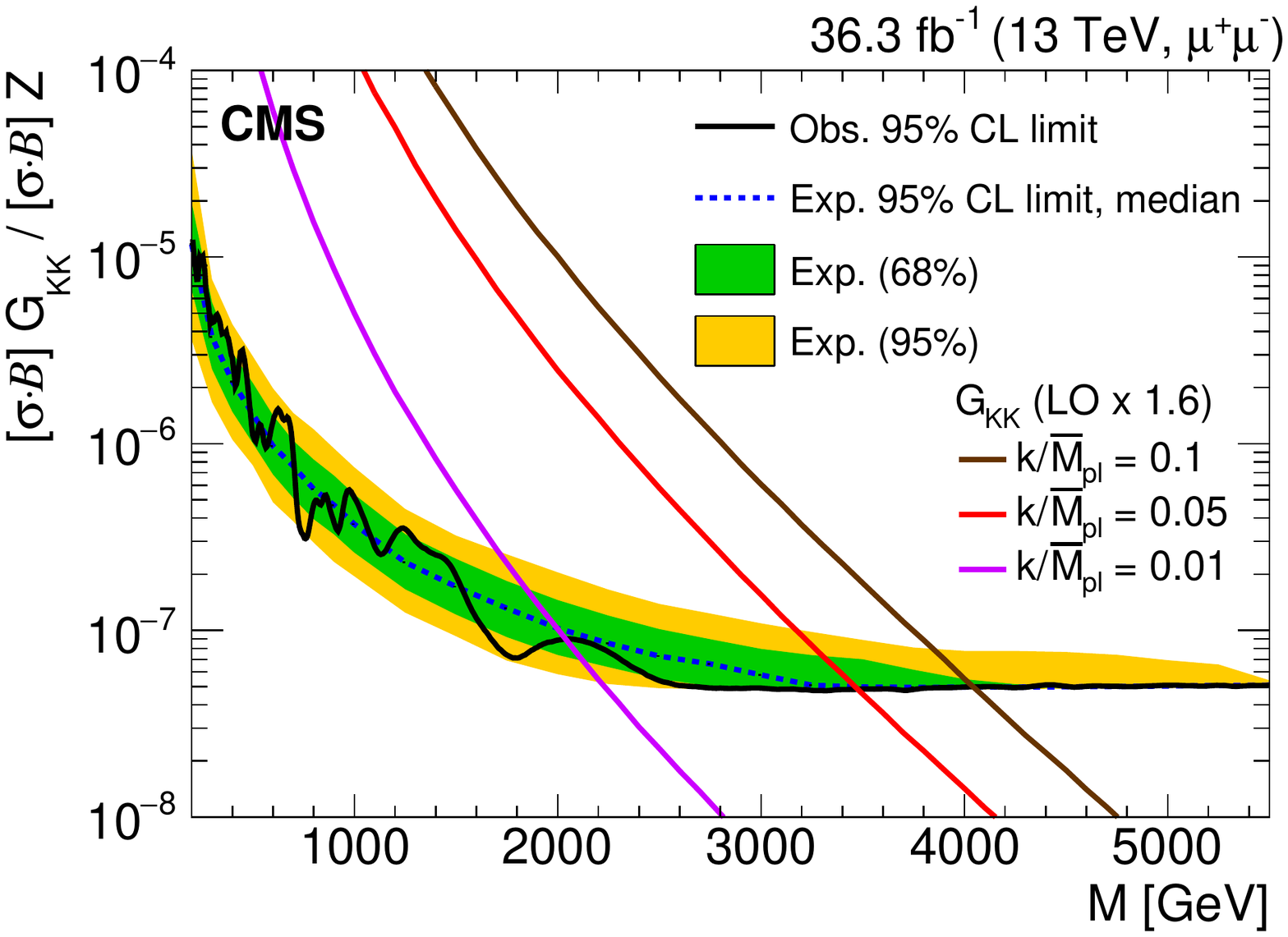}
\includegraphics[width=0.49\textwidth]{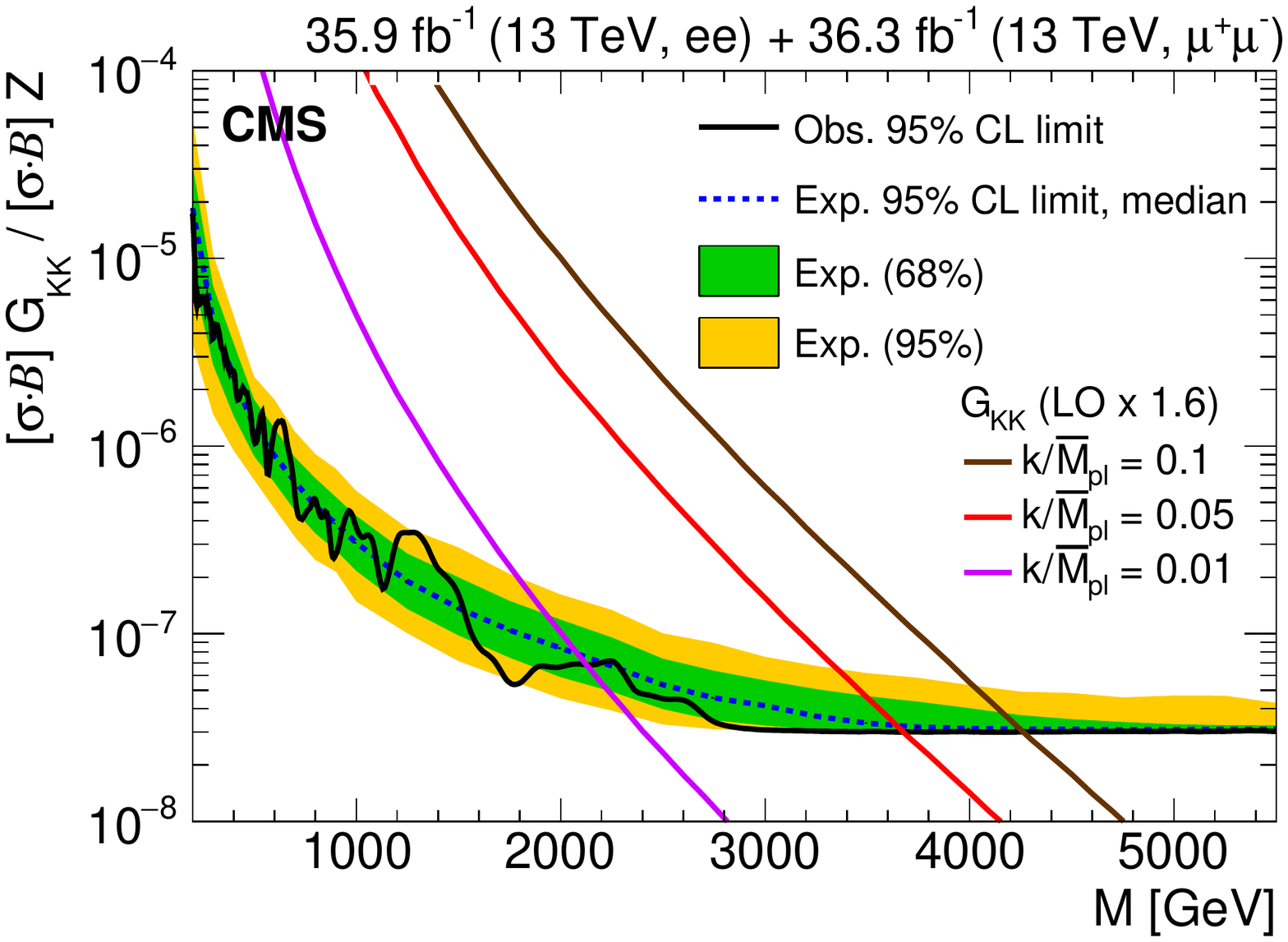}
\caption{
The upper limits at 95\% \CL on the product of production cross section and branching fraction for a spin-2 resonance, relative to the product of production cross section and branching fraction of a \PZ boson, for the dielectron channel (left), dimuon channel (right), and their combination (lower).
The shaded bands correspond to the 68 and 95\% quantiles for the expected limits.
Theoretical predictions for the spin-2 resonances with widths equal to 0.01, 0.36, and 1.42\GeV corresponding to coupling parameters $k/\overline{M}_\mathrm{Pl}$ of 0.01, 0.05, and 0.10 are shown for comparison.
}
\label{fig:limitsRS}
\end{figure}

The expected and observed limits for a spin-2 resonance with intrinsic widths of 0.01, 0.36, and 1.42\GeV corresponding to coupling parameters $k/\overline{M}_\mathrm{Pl}$ of 0.01, 0.05, and 0.10, are shown in Fig.~\ref{fig:limitsRS} for the dielectron channel, dimuon channel, and their combination.
Table~\ref{tab:massLimitsSpin2} presents the values of the observed and expected 95\% \CL lower limits of the aforementioned models.
The signal production cross sections, calculated using the \PYTHIA~8.2 program with the NNPDF2.3 PDFs at LO, are multiplied by a $K$-factor of 1.6 to account for NLO effects~\cite{Mathews:2005bw}.
The PI contribution to the production cross sections is small enough to be ignored.

\begin{table}[!hbt]
\centering
\topcaption{
The observed and expected 95\% \CL lower limits on the masses of spin-2 resonances with widths equal to 0.01, 0.36 and 1.42\GeV corresponding to coupling parameters $k/\overline{M}_\mathrm{Pl}$ of 0.01, 0.05, and 0.10.
}
\begin{tabular}{ccccccc}
\multirow{2}{*}{Channel}  & \multicolumn{2}{c}{$k/\overline{M}_\mathrm{Pl} = 0.01$} & \multicolumn{2}{c}{$k/\overline{M}_\mathrm{Pl} = 0.05$}  & \multicolumn{2}{c}{$k/\overline{M}_\mathrm{Pl} = 0.1$}\\
                          & Obs. [\TeVns{}] & Exp. [\TeVns{}]      & Obs. [\TeVns{}]  & Exp. [\TeVns{}]   & Obs. [\TeVns{}]  & Exp. [\TeVns{}]      \\\hline
\Pe\Pe                    &  1.85      & 1.85            & 3.30        & 3.30   & 3.90  & 3.90        \\
$\mu^+\mu^-$              &  2.05      & 2.00            & 3.50        & 3.50   & 4.05  & 4.05        \\
\Pe\Pe ~+ $\mu^+\mu^-$    &  2.10      & 2.05            & 3.65        & 3.60   & 4.25  & 4.25        \\
\end{tabular}
\label{tab:massLimitsSpin2}
\end{table}

\begin{figure*} [htb]
\centering
\includegraphics[width=0.49\textwidth]{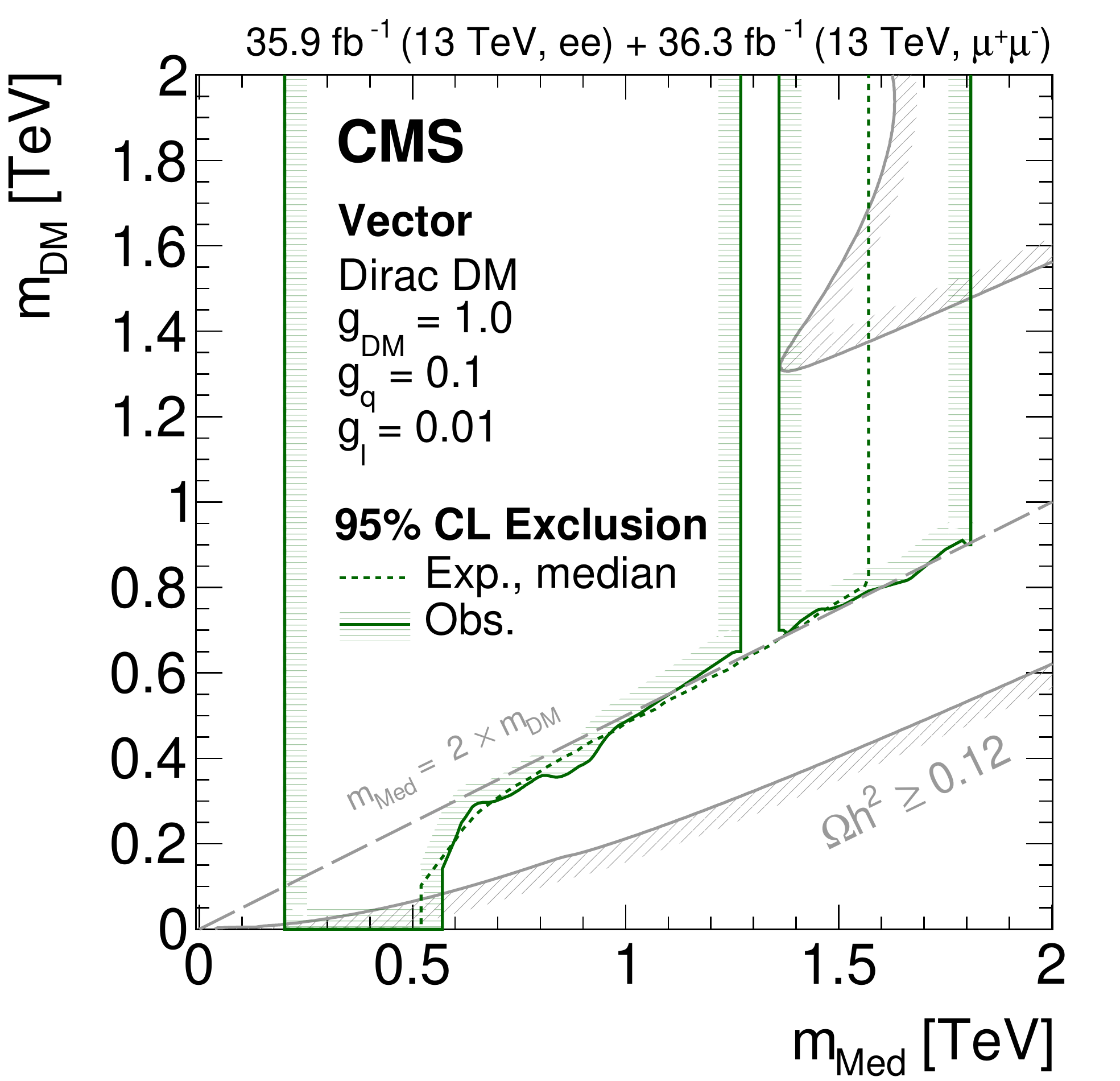} \hfill
\includegraphics[width=0.49\textwidth]{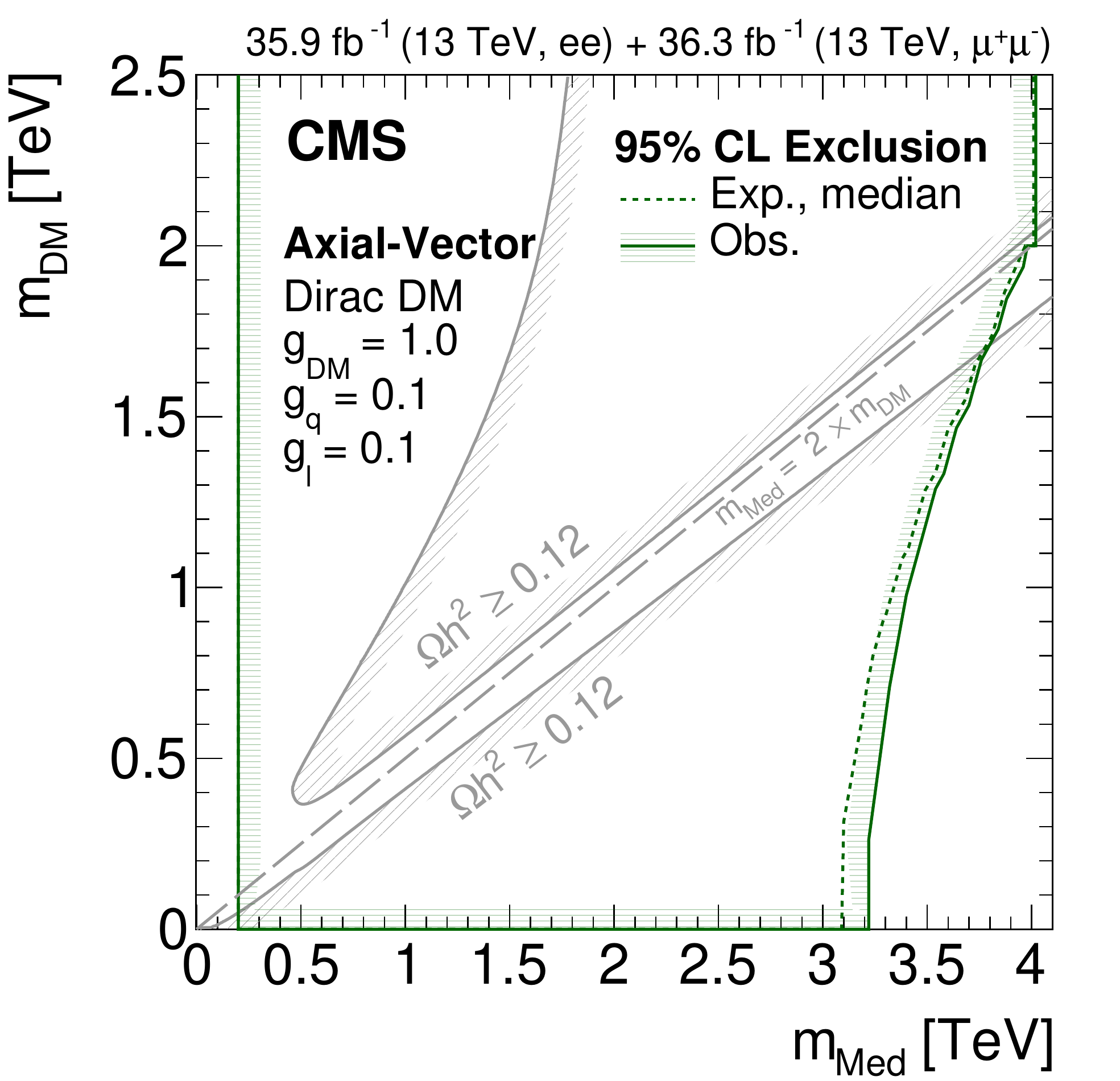}
\caption{
Limits at 95\% confidence level for the masses of the DM particle, which is assumed to be Dirac fermion, and its associated mediator, in a simplified model of DM production via a vector (left) or axial vector (right) mediator.
The parameter exclusion is obtained by comparing the limits on product of the production cross section and the branching fraction for decay to a \PZ boson, with the values obtained from calculations in the simplified model.
For each combination of the DM particle and mediator mass values, the width of the mediator is taken into account in the limit calculation.
The lines with the hatching represents the excluded regions.
The solid grey lines, marked as ``$\Omega h^{2} \ge 0.12$'', correspond to parameter regions that reproduce the observed DM relic density in the universe~\cite{Backovic:2015tpt,Backovic:2013dpa,Ade:2015xua,Albert:2017onk}, with the hatched area indicating the region where the DM relic abundance exceeds the observed value.
}
\label{fig:DM}
\end{figure*}

The results are also interpreted in the context of a simplified model with a DM particle that has sizeable interactions with SM fermions through an additional spin-1 high-mass particle mediating the SM-DM interaction.
In the simplified model under consideration~\cite{Albert:2017onk}, only one DM particle exists, which is assumed to be a Dirac fermion.
Limits are presented in Fig.~\ref{fig:DM} for two cases with different sets of benchmark coupling values.
The first case corresponds to a vector mediator with small couplings to leptons while the second one corresponds to an axial-vector mediator with equal couplings to quarks and leptons.
The cross sections for lepton production are calculated at NLO in QCD, using the \textsc{DMsimp} implementation~\cite{Backovic:2015soa} of the simplified model in \MGvATNLO version 2.5.2~\cite{Alwall:2014hca}.
Assuming the optimistic axial-vector coupling scenario and $m_\text{DM} > m_{\text{Med}}/2$, the signal cross section for the production of an electron or muon pair within the analysis acceptance ranges between approximately 100\unit{pb} at low values of the mediator mass (around 200\GeV), and 0.1\unit{fb} for higher values (around 4\TeV).
The partial and total mediator decay widths, calculated at LO in QCD, are included via the \textsc{MadWidth} package~\cite{Alwall:2014bza}.

While the DM particle is not probed directly, its mass indirectly modulates the sensitivity of the dilepton search.
For low values of the DM particle mass $m_\text{DM} < m_\text{Med}/2$, the mediator boson will dominantly decay into DM particles, thus reducing the branching fraction to leptons, and making the mediator harder to probe in this search.
At high values of the DM particle mass $m_\text{DM} > m_\text{Med}/2$, the mediator cannot decay to the DM particles and the leptonic branching fraction becomes sizeable.
In the vector mediator model, the relatively small leptonic couplings mostly limit the sensitivity of this analysis to the regime of $m_\text{DM} > m_\text{Med}/2$.
This regime is especially interesting to probe since it is almost inaccessible to typical searches based on missing transverse momentum~\cite{Sirunyan:2017hci}.
In the axial-vector mediator model, the leptonic couplings of the mediator are sizeable and an exclusion is also possible for $m_\text{DM} < m_\text{Med}/2$.
In the former case, the limit on the mediator mass reaches up to 1.8\TeV, depending on the mass of the DM particle; while in the latter case the limit reaches the  3.0--4.0\TeV, depending again on the mass of the DM particle.
In the vector mediator model, the observed exclusion reaches up to values of the mediator mass equal to approximately 1.8 (0.6)\TeV above (below) the diagonal, $m_\text{DM} = m_{\text{Med}}/2$.
In the region where the mediator mass is equal to approximately 1.3\TeV, an upward fluctuation in the data (Fig.~\ref{fig:limits2}) results in a small above-diagonal region that is not excluded.
Assuming that there is no new physics other than the mediator and DM particle, the relic density of DM in the universe, $\Omega h^{2}$, can be calculated.
Regions of parameter space that reproduce the observed value $\Omega h^{2} \approx 0.12$~\cite{Ade:2015xua}, are indicated in Fig.~\ref{fig:DM}.
Numerical values are obtained from Ref.~\cite{Albert:2017onk}, where they were calculated using \textsc{MadDM} in version 2.06~\cite{Backovic:2013dpa,Backovic:2015tpt}.
In the hatched area, the DM would be overabundant in the universe.
However, it is unlikely that new physics is fully described by the simplified model considered here, and the relic density constraint is not a stringent constraint, as additional new phenomena may modify its calculated value.

\clearpage
\section{Summary}
\label{sec:summary}

A search for narrow resonances in dielectron and dimuon invariant mass spectra has been performed using data recorded in 2016 from proton-proton collisions at $\sqrt{s}=13\TeV$.
The integrated luminosity for the dielectron sample is \anaLumiee\fbinv and for the dimuon sample is \anaLumimumu\fbinv.
Observations are in agreement with standard model expectations.
Upper limits at 95\% confidence level on the product of a narrow-resonance production cross section and branching fraction to dileptons have been calculated in a model-independent manner to enable interpretation in the framework of models predicting a narrow dielectron or dimuon resonance.
A scan of different intrinsic width hypotheses is performed.

Limits are set on the masses of various hypothetical particles.
For the $\ZPSSM$ particle, which arises in the sequential standard model, and for the superstring-inspired $\ZPPSI$ particle, 95\% confidence level lower mass limits for the combined channels are found to be $\limitZssm$ and $\limitZpsi\TeV$, respectively.
These limits extend the previous ones from CMS by 1.1\TeV in both models.
The corresponding limits for Kaluza--Klein gravitons arising in the Randall--Sundrum model of extra dimensions with coupling parameters $k/\overline{M}_\mathrm{Pl}$ of 0.01, 0.05, and 0.10 are $\limitGone$, $\limitGtwo$, and $\limitGthree\TeV$, respectively.
The limits extend previous published CMS results by 0.6 (1.1)\TeV for a $k/\overline{M}_\mathrm{Pl}$ value of 0.01 (0.10).
Finally, limits at 95\% confidence level are obtained for the masses of the dark matter particle and its associated mediator, in a simplified model of dark matter production via a vector or axial vector mediator.

\begin{acknowledgments}

We congratulate our colleagues in the CERN accelerator departments for the excellent performance of the LHC and thank the technical and administrative staffs at CERN and at other CMS institutes for their contributions to the success of the CMS effort. In addition, we gratefully acknowledge the computing centres and personnel of the Worldwide LHC Computing Grid for delivering so effectively the computing infrastructure essential to our analyses. Finally, we acknowledge the enduring support for the construction and operation of the LHC and the CMS detector provided by the following funding agencies: BMWFW and FWF (Austria); FNRS and FWO (Belgium); CNPq, CAPES, FAPERJ, and FAPESP (Brazil); MES (Bulgaria); CERN; CAS, MoST, and NSFC (China); COLCIENCIAS (Colombia); MSES and CSF (Croatia); RPF (Cyprus); SENESCYT (Ecuador); MoER, ERC IUT, and ERDF (Estonia); Academy of Finland, MEC, and HIP (Finland); CEA and CNRS/IN2P3 (France); BMBF, DFG, and HGF (Germany); GSRT (Greece); NKFIA (Hungary); DAE and DST (India); IPM (Iran); SFI (Ireland); INFN (Italy); MSIP and NRF (Republic of Korea); LAS (Lithuania); MOE and UM (Malaysia); BUAP, CINVESTAV, CONACYT, LNS, SEP, and UASLP-FAI (Mexico); MBIE (New Zealand); PAEC (Pakistan); MSHE and NSC (Poland); FCT (Portugal); JINR (Dubna); MON, RosAtom, RAS, RFBR and RAEP (Russia); MESTD (Serbia); SEIDI, CPAN, PCTI and FEDER (Spain); Swiss Funding Agencies (Switzerland); MST (Taipei); ThEPCenter, IPST, STAR, and NSTDA (Thailand); TUBITAK and TAEK (Turkey); NASU and SFFR (Ukraine); STFC (United Kingdom); DOE and NSF (USA).

\hyphenation{Rachada-pisek} Individuals have received support from the Marie-Curie programme and the European Research Council and Horizon 2020 Grant, contract No. 675440 (European Union); the Leventis Foundation; the A. P. Sloan Foundation; the Alexander von Humboldt Foundation; the Belgian Federal Science Policy Office; the Fonds pour la Formation \`a la Recherche dans l'Industrie et dans l'Agriculture (FRIA-Belgium); the Agentschap voor Innovatie door Wetenschap en Technologie (IWT-Belgium); the F.R.S.-FNRS and FWO (Belgium) under the ``Excellence of Science - EOS" - be.h project n. 30820817; the Ministry of Education, Youth and Sports (MEYS) of the Czech Republic; the Lend\"ulet ("Momentum") Programme and the J\'anos Bolyai Research Scholarship of the Hungarian Academy of Sciences, the New National Excellence Program \'UNKP, the NKFIA research grants 123842, 123959, 124845, 124850 and 125105 (Hungary); the Council of Science and Industrial Research, India; the HOMING PLUS programme of the Foundation for Polish Science, cofinanced from European Union, Regional Development Fund, the Mobility Plus programme of the Ministry of Science and Higher Education, the National Science Center (Poland), contracts Harmonia 2014/14/M/ST2/00428, Opus 2014/13/B/ST2/02543, 2014/15/B/ST2/03998, and 2015/19/B/ST2/02861, Sonata-bis 2012/07/E/ST2/01406; the National Priorities Research Program by Qatar National Research Fund; the Programa Estatal de Fomento de la Investigaci{\'o}n Cient{\'i}fica y T{\'e}cnica de Excelencia Mar\'{\i}a de Maeztu, grant MDM-2015-0509 and the Programa Severo Ochoa del Principado de Asturias; the Thalis and Aristeia programmes cofinanced by EU-ESF and the Greek NSRF; the Rachadapisek Sompot Fund for Postdoctoral Fellowship, Chulalongkorn University and the Chulalongkorn Academic into Its 2nd Century Project Advancement Project (Thailand); the Welch Foundation, contract C-1845; and the Weston Havens Foundation (USA).

\end{acknowledgments}

\bibliography{auto_generated}

\providecommand{\href}[2]{#2}\begingroup\raggedright\begin{thebibliography}{10}%
\makeatletter
\providecommand{\hrefCMSnoop }[0]{\@secondoftwo}%
\makeatother
\providecommand{\doi}{\texttt{doi:}\begingroup \urlstyle{tt}\Url}

\bibitem{Leike:1998wr}
\hrefCMSnoop {}{A.~Leike, ``The phenomenology of extra neutral gauge bosons'',}
  \textit{ Phys. Rept.} \textbf{ 317} (1999) 143,
  \href{http://dx.doi.org/10.1016/S0370-1573(98)00133-1}{\doi{10.1016/S0370-1573(98)00133-1}},
\href{http://www.arXiv.org/abs/hep-ph/9805494}{\texttt{arXiv:hep-ph/9805494}}.

\bibitem{Zp_SSM_1}
\hrefCMSnoop {}{P.~Langacker, ``The physics of heavy {$\zp$} gauge bosons'',}
  \textit{ Rev. Mod. Phys.} \textbf{ 81} (2009) 1199,
  \href{http://dx.doi.org/10.1103/RevModPhys.81.1199}{\doi{10.1103/RevModPhys.81.1199}},
\href{http://www.arXiv.org/abs/0801.1345}{\texttt{arXiv:0801.1345}}.

\bibitem{Randall:1999ee}
\hrefCMSnoop {}{L.~Randall and R.~Sundrum, ``A large mass hierarchy from a
  small extra dimension'',} \textit{ Phys. Rev. Lett.} \textbf{ 83} (1999)
  3370,
  \href{http://dx.doi.org/10.1103/PhysRevLett.83.3370}{\doi{10.1103/PhysRevLett.83.3370}},
\href{http://www.arXiv.org/abs/hep-ph/9905221}{\texttt{arXiv:hep-ph/9905221}}.

\bibitem{Ade:2015xua}
\hrefCMSnoop {}{{Planck} Collaboration, ``{Planck 2015 results. XIII.
  Cosmological parameters}'',} \textit{ Astron. Astrophys.} \textbf{ 594}
  (2016) A13,
  \href{http://dx.doi.org/10.1051/0004-6361/201525830}{\doi{10.1051/0004-6361/201525830}},
\href{http://www.arXiv.org/abs/1502.01589}{\texttt{arXiv:1502.01589}}.

\bibitem{Carena:2004xs}
\hrefCMSnoop {}{M.~S. Carena, A.~Daleo, B.~A. Dobrescu, and T.~M.~P. Tait,
  ``$\zp$ gauge bosons at the {T}evatron'',} \textit{ Phys. Rev. D} \textbf{
  70} (2004) 093009,
  \href{http://dx.doi.org/10.1103/PhysRevD.70.093009}{\doi{10.1103/PhysRevD.70.093009}},
\href{http://www.arXiv.org/abs/hep-ph/0408098}{\texttt{arXiv:hep-ph/0408098}}.

\bibitem{Accomando:2010fz}
E.~Accomando\hrefCMSnoop {}{ {et~al.}, ``{$\zp$} physics with early {LHC}
  data'',} \textit{ Phys. Rev. D} \textbf{ 83} (2011) 075012,
  \href{http://dx.doi.org/10.1103/PhysRevD.83.075012}{\doi{10.1103/PhysRevD.83.075012}},
  \href{http://www.arXiv.org/abs/1010.6058}{\texttt{arXiv:1010.6058}}.

\bibitem{Altar:1989}
\hrefCMSnoop {}{G.~Altarelli, B.~Mele, and M.~Ruiz-Altaba, ``Searching for new
  heavy vector bosons in {$\Pp\Pap$} colliders'',} \textit{ Z. Phys. C}
  \textbf{ 45} (1989) 109,
  \href{http://dx.doi.org/10.1007/BF01556677}{\doi{10.1007/BF01556677}}.
  [Erratum: Z. Phys. C 47 (1990) 676].

\bibitem{Zp_PSI_3}
\hrefCMSnoop {}{J.~L. Hewett and T.~G. Rizzo, ``Low-energy phenomenology of
  superstring-inspired {E}$_6$ models'',} \textit{ Phys. Rept.} \textbf{ 183}
  (1989) 193,
  \href{http://dx.doi.org/10.1016/0370-1573(89)90071-9}{\doi{10.1016/0370-1573(89)90071-9}}.

\bibitem{Accomando:2013sfa}
E.~Accomando\hrefCMSnoop {}{ {et~al.}, ``{Z' at the LHC: Interference and
  Finite Width Effects in Drell-Yan}'',} \textit{ JHEP} \textbf{ 10} (2013)
  153,
  \href{http://dx.doi.org/10.1007/JHEP10(2013)153}{\doi{10.1007/JHEP10(2013)153}},
\href{http://www.arXiv.org/abs/1304.6700}{\texttt{arXiv:1304.6700}}.

\bibitem{Chatrchyan:2011wq}
\hrefCMSnoop {}{{CMS Collaboration}, ``Search for resonances in the dilepton
  mass distribution in {$\Pp\Pp$} collisions at $\sqrt{s}=7$ {TeV}'',} \textit{
  JHEP} \textbf{ 05} (2011) 093,
  \href{http://dx.doi.org/10.1007/JHEP05(2011)093}{\doi{10.1007/JHEP05(2011)093}},
  \href{http://www.arXiv.org/abs/1103.0981}{\texttt{arXiv:1103.0981}}.

\bibitem{Chatrchyan:2012it}
\hrefCMSnoop {}{{CMS Collaboration}, ``{Search for narrow resonances in
  dilepton mass spectra in {$\Pp\Pp$} collisions at $\sqrt{s}=7$ TeV}'',}
  \textit{ Phys. Lett. B} \textbf{ 714} (2012) 158,
  \href{http://dx.doi.org/10.1016/j.physletb.2012.06.051}{\doi{10.1016/j.physletb.2012.06.051}},
\href{http://www.arXiv.org/abs/1206.1849}{\texttt{arXiv:1206.1849}}.

\bibitem{Chatrchyan:2012oaa}
\hrefCMSnoop {}{{CMS Collaboration}, ``{Search for heavy narrow dilepton
  resonances in {$\Pp\Pp$} collisions at $\sqrt{s}=7$ TeV and $\sqrt{s}=8$
  TeV}'',} \textit{ Phys. Lett. B} \textbf{ 720} (2013) 63,
  \href{http://dx.doi.org/10.1016/j.physletb.2013.02.003}{\doi{10.1016/j.physletb.2013.02.003}},
\href{http://www.arXiv.org/abs/1212.6175}{\texttt{arXiv:1212.6175}}.

\bibitem{Khachatryan:2014fba}
\hrefCMSnoop {}{{CMS Collaboration}, ``{Search for physics beyond the standard
  model in dilepton mass spectra in proton-proton collisions at $ \sqrt{s}=8 $
  TeV}'',} \textit{ JHEP} \textbf{ 04} (2015) 025,
  \href{http://dx.doi.org/10.1007/JHEP04(2015)025}{\doi{10.1007/JHEP04(2015)025}},
\href{http://www.arXiv.org/abs/1412.6302}{\texttt{arXiv:1412.6302}}.

\bibitem{Khachatryan:2016zqb}
\hrefCMSnoop {}{{CMS Collaboration}, ``{Search for narrow resonances in
  dilepton mass spectra in proton-proton collisions at $\sqrt{s}$ = 13 TeV and
  combination with 8 TeV data}'',} \textit{ Phys. Lett. B} \textbf{ 768} (2017)
  57,
  \href{http://dx.doi.org/10.1016/j.physletb.2017.02.010}{\doi{10.1016/j.physletb.2017.02.010}},
\href{http://www.arXiv.org/abs/1609.05391}{\texttt{arXiv:1609.05391}}.

\bibitem{Aad:2011xp}
\hrefCMSnoop {}{{ATLAS Collaboration}, ``{Search for high mass dilepton
  resonances in {$\Pp\Pp$} collisions at $\sqrt{s}=7$ TeV with the ATLAS
  experiment}'',} \textit{ Phys. Lett. B} \textbf{ 700} (2011) 163,
  \href{http://dx.doi.org/10.1016/j.physletb.2011.04.044}{\doi{10.1016/j.physletb.2011.04.044}},
\href{http://www.arXiv.org/abs/1103.6218}{\texttt{arXiv:1103.6218}}.

\bibitem{Aad:2012hf}
\hrefCMSnoop {}{{ATLAS Collaboration}, ``{Search for high-mass resonances
  decaying to dilepton final states in {$\Pp\Pp$} collisions at
  $\sqrt{s}=7\TeV$ with the ATLAS detector}'',} \textit{ JHEP} \textbf{ 11}
  (2012) 138,
  \href{http://dx.doi.org/10.1007/JHEP11(2012)138}{\doi{10.1007/JHEP11(2012)138}},
\href{http://www.arXiv.org/abs/1209.2535}{\texttt{arXiv:1209.2535}}.

\bibitem{Aad:2014cka}
\hrefCMSnoop {}{{ATLAS Collaboration}, ``{Search for high-mass dilepton
  resonances in pp collisions at $\sqrt{s}=8$~TeV with the ATLAS detector}'',}
  \textit{ Phys. Rev. D} \textbf{ 90} (2014) 052005,
  \href{http://dx.doi.org/10.1103/PhysRevD.90.052005}{\doi{10.1103/PhysRevD.90.052005}},
\href{http://www.arXiv.org/abs/1405.4123}{\texttt{arXiv:1405.4123}}.

\bibitem{Aaboud:2017buh}
\hrefCMSnoop {}{{ATLAS Collaboration}, ``{Search for new high-mass phenomena in
  the dilepton final state using 36.1 fb$^{-1}$ of proton-proton collision data
  at $\sqrt{s}$ = 13 TeV with the ATLAS detector}'',} \textit{ JHEP} \textbf{
  10} (2017) 182,
  \href{http://dx.doi.org/10.1007/JHEP10(2017)182}{\doi{10.1007/JHEP10(2017)182}},
\href{http://www.arXiv.org/abs/1707.02424}{\texttt{arXiv:1707.02424}}.

\bibitem{Randall:1999vf}
\hrefCMSnoop {}{L.~Randall and R.~Sundrum, ``An alternative to
  compactification'',} \textit{ Phys. Rev. Lett.} \textbf{ 83} (1999) 4690,
  \href{http://dx.doi.org/10.1103/PhysRevLett.83.4690}{\doi{10.1103/PhysRevLett.83.4690}},
\href{http://www.arXiv.org/abs/hep-th/9906064}{\texttt{arXiv:hep-th/9906064}}.

\bibitem{CDF_Zp}
\hrefCMSnoop {}{{CDF} Collaboration, ``Search for high-mass {$\EE$} resonances
  in {$\Pp\Pap$} collisions at $\sqrt{s} = 1.96$ {TeV}'',} \textit{ Phys. Rev.
  Lett.} \textbf{ 102} (2009) 031801,
  \href{http://dx.doi.org/10.1103/PhysRevLett.102.031801}{\doi{10.1103/PhysRevLett.102.031801}},
\href{http://www.arXiv.org/abs/0810.2059}{\texttt{arXiv:0810.2059}}.

\bibitem{CDF_RS}
\hrefCMSnoop {}{{CDF} Collaboration, ``A search for high-mass resonances
  decaying to dimuons at {CDF}'',} \textit{ Phys. Rev. Lett.} \textbf{ 102}
  (2009) 091805,
  \href{http://dx.doi.org/10.1103/PhysRevLett.102.091805}{\doi{10.1103/PhysRevLett.102.091805}},
\href{http://www.arXiv.org/abs/0811.0053}{\texttt{arXiv:0811.0053}}.

\bibitem{D0_RS}
\hrefCMSnoop {}{{D0} Collaboration, ``Search for {R}andall--{S}undrum gravitons
  in the dielectron and diphoton final states with 5.4 fb$^{-1}$ of data from
  {$\Pp\Pap$} collisions at $\sqrt{s} = 1.96$ {TeV}'',} \textit{ Phys. Rev.
  Lett.} \textbf{ 104} (2010) 241802,
  \href{http://dx.doi.org/10.1103/PhysRevLett.104.241802}{\doi{10.1103/PhysRevLett.104.241802}},
\href{http://www.arXiv.org/abs/1004.1826}{\texttt{arXiv:1004.1826}}.

\bibitem{D0_Zp}
\hrefCMSnoop {}{{D0} Collaboration, ``Search for a heavy neutral gauge boson in
  the dielectron channel with 5.4 fb$^{-1}$ of {$\Pp\Pap$} collisions at
  $\sqrt{s} = 1.96$ {TeV}'',} \textit{ Phys. Lett. B} \textbf{ 695} (2011) 88,
  \href{http://dx.doi.org/10.1016/j.physletb.2010.10.059}{\doi{10.1016/j.physletb.2010.10.059}},
\href{http://www.arXiv.org/abs/1008.2023}{\texttt{arXiv:1008.2023}}.

\bibitem{CDF_SSM}
\hrefCMSnoop {}{{CDF} Collaboration, ``Search for high mass resonances decaying
  to muon pairs in $\sqrt{s} = 1.96$ {TeV} {$\Pp\Pap$} collisions'',} \textit{
  Phys. Rev. Lett.} \textbf{ 106} (2011) 121801,
  \href{http://dx.doi.org/10.1103/PhysRevLett.106.121801}{\doi{10.1103/PhysRevLett.106.121801}},
  \href{http://www.arXiv.org/abs/1101.4578}{\texttt{arXiv:1101.4578}}.

\bibitem{CDF_RSele}
\hrefCMSnoop {}{{CDF} Collaboration, ``Search for new dielectron resonances and
  {R}andall--{S}undrum gravitons at the {C}ollider {D}etector at {F}ermilab'',}
  \textit{ Phys. Rev. Lett.} \textbf{ 107} (2011) 051801,
  \href{http://dx.doi.org/10.1103/PhysRevLett.107.051801}{\doi{10.1103/PhysRevLett.107.051801}},
  \href{http://www.arXiv.org/abs/1103.4650}{\texttt{arXiv:1103.4650}}.

\bibitem{Albert:2017onk}
\hrefCMSnoop {}{A.~Albert {et~al.}, ``{Recommendations of the LHC Dark Matter
  Working Group: Comparing LHC searches for heavy mediators of dark matter
  production in visible and invisible decay channels}'',} (2017).
\href{http://www.arXiv.org/abs/1703.05703}{\texttt{arXiv:1703.05703}}.

\bibitem{Backovic:2015soa}
M.~Backovi\'{c}\hrefCMSnoop {}{ {et~al.}, ``{Higher-order QCD predictions for
  dark matter production at the LHC in simplified models with $s$ channel
  mediators}'',} \textit{ Eur. Phys. J. C} \textbf{ 75} (2015) 482,
  \href{http://dx.doi.org/10.1140/epjc/s10052-015-3700-6}{\doi{10.1140/epjc/s10052-015-3700-6}},
\href{http://www.arXiv.org/abs/1508.05327}{\texttt{arXiv:1508.05327}}.

\bibitem{1748-0221-12-01-C01048}
\hrefCMSnoop {}{C.~Battilana, ``{The CMS muon system: status and upgrades for
  LHC Run-2 and performance of muon reconstruction with 13~TeV data}'',}
  \textit{ JINST}
  \href{http://dx.doi.org/10.1088/1748-0221/12/01/C01048}{\doi{10.1088/1748-0221/12/01/C01048}}.

\bibitem{Chatrchyan:2008zzk}
\hrefCMSnoop {}{{CMS Collaboration}, ``The {CMS} experiment at the {CERN}
  {LHC}'',} \textit{ JINST} \textbf{ 3} (2008) S08004,
\href{http://dx.doi.org/10.1088/1748-0221/3/08/S08004}{\doi{10.1088/1748-0221/3/08/S08004}}.

\bibitem{Khachatryan:2016bia}
\hrefCMSnoop {}{{CMS Collaboration}, ``{The CMS trigger system}'',} \textit{
  JINST} \textbf{ 12} (2017) P01020,
  \href{http://dx.doi.org/10.1088/1748-0221/12/01/P01020}{\doi{10.1088/1748-0221/12/01/P01020}},
\href{http://www.arXiv.org/abs/1609.02366}{\texttt{arXiv:1609.02366}}.

\bibitem{Nason:2004rx}
\hrefCMSnoop {}{P.~Nason, ``A new method for combining {NLO} {QCD} with shower
  {M}onte {C}arlo algorithms'',} \textit{ JHEP} \textbf{ 11} (2004) 040,
  \href{http://dx.doi.org/10.1088/1126-6708/2004/11/040}{\doi{10.1088/1126-6708/2004/11/040}},
\href{http://www.arXiv.org/abs/hep-ph/0409146}{\texttt{arXiv:hep-ph/0409146}}.

\bibitem{Frixione:2007vw}
\hrefCMSnoop {}{S.~Frixione, P.~Nason, and C.~Oleari, ``Matching {NLO} {QCD}
  computations with parton shower simulations: the {POWHEG} method'',} \textit{
  JHEP} \textbf{ 11} (2007) 070,
  \href{http://dx.doi.org/10.1088/1126-6708/2007/11/070}{\doi{10.1088/1126-6708/2007/11/070}},
\href{http://www.arXiv.org/abs/0709.2092}{\texttt{arXiv:0709.2092}}.

\bibitem{Alioli:2010xd}
\hrefCMSnoop {}{S.~Alioli, P.~Nason, C.~Oleari, and E.~Re, ``A general
  framework for implementing {NLO} calculations in shower {M}onte {C}arlo
  programs: the {POWHEG} {BOX}'',} \textit{ JHEP} \textbf{ 06} (2010) 043,
  \href{http://dx.doi.org/10.1007/JHEP06(2010)043}{\doi{10.1007/JHEP06(2010)043}},
\href{http://www.arXiv.org/abs/1002.2581}{\texttt{arXiv:1002.2581}}.

\bibitem{Alioli:2008gx}
\hrefCMSnoop {}{S.~Alioli, P.~Nason, C.~Oleari, and E.~Re, ``{NLO vector-boson
  production matched with shower in POWHEG}'',} \textit{ JHEP} \textbf{ 07}
  (2008) 060,
  \href{http://dx.doi.org/10.1088/1126-6708/2008/07/060}{\doi{10.1088/1126-6708/2008/07/060}},
\href{http://www.arXiv.org/abs/0805.4802}{\texttt{arXiv:0805.4802}}.

\bibitem{Frixione:2007nw}
\hrefCMSnoop {}{S.~Frixione, P.~Nason, and G.~Ridolfi, ``{A positive-weight
  next-to-leading-order Monte Carlo for heavy flavour hadroproduction}'',}
  \textit{ JHEP} \textbf{ 09} (2007) 126,
  \href{http://dx.doi.org/10.1088/1126-6708/2007/09/126}{\doi{10.1088/1126-6708/2007/09/126}},
\href{http://www.arXiv.org/abs/0707.3088}{\texttt{arXiv:0707.3088}}.

\bibitem{Re:2010bp}
\hrefCMSnoop {}{E.~Re, ``Single-top {$\PW\cPqt$}-channel production matched
  with parton showers using the {POWHEG} method'',} \textit{ Eur. Phys. J. C}
  \textbf{ 71} (2011) 1547,
  \href{http://dx.doi.org/10.1140/epjc/s10052-011-1547-z}{\doi{10.1140/epjc/s10052-011-1547-z}},
\href{http://www.arXiv.org/abs/1009.2450}{\texttt{arXiv:1009.2450}}.

\bibitem{Ball:2014uwa}
\hrefCMSnoop {}{{NNPDF} Collaboration, ``{Parton distributions for the LHC Run
  II}'',} \textit{ JHEP} \textbf{ 04} (2015) 040,
  \href{http://dx.doi.org/10.1007/JHEP04(2015)040}{\doi{10.1007/JHEP04(2015)040}},
\href{http://www.arXiv.org/abs/1410.8849}{\texttt{arXiv:1410.8849}}.

\bibitem{Sjostrand:2014zea}
T.~Sj{\"o}strand\hrefCMSnoop {}{ {et~al.}, ``{An Introduction to PYTHIA
  8.2}'',} \textit{ Comput. Phys. Commun.} \textbf{ 191} (2015) 159,
  \href{http://dx.doi.org/10.1016/j.cpc.2015.01.024}{\doi{10.1016/j.cpc.2015.01.024}},
\href{http://www.arXiv.org/abs/1410.3012}{\texttt{arXiv:1410.3012}}.

\bibitem{Khachatryan:2015pea}
\hrefCMSnoop {}{{CMS Collaboration}, ``{Event generator tunes obtained from
  underlying event and multiparton scattering measurements}'',} \textit{ Eur.
  Phys. J. C} \textbf{ 76} (2016) 155,
  \href{http://dx.doi.org/10.1140/epjc/s10052-016-3988-x}{\doi{10.1140/epjc/s10052-016-3988-x}},
\href{http://www.arXiv.org/abs/1512.00815}{\texttt{arXiv:1512.00815}}.

\bibitem{Li:2012wna}
\hrefCMSnoop {}{Y.~Li and F.~Petriello, ``Combining {QCD} and electroweak
  corrections to dilepton production in {FEWZ}'',} \textit{ Phys. Rev. D}
  \textbf{ 86} (2012) 094034,
  \href{http://dx.doi.org/10.1103/PhysRevD.86.094034}{\doi{10.1103/PhysRevD.86.094034}},
  \href{http://www.arXiv.org/abs/1208.5967}{\texttt{arXiv:1208.5967}}.

\bibitem{Manohar:2016nzj}
\hrefCMSnoop {}{A.~Manohar, P.~Nason, G.~P. Salam, and G.~Zanderighi, ``{How
  bright is the proton? A precise determination of the photon parton
  distribution function}'',} \textit{ Phys. Rev. Lett.} \textbf{ 117} (2016)
  242002,
  \href{http://dx.doi.org/10.1103/PhysRevLett.117.242002}{\doi{10.1103/PhysRevLett.117.242002}},
\href{http://www.arXiv.org/abs/1607.04266}{\texttt{arXiv:1607.04266}}.

\bibitem{Butterworth:2015oua}
\hrefCMSnoop {}{J.~Butterworth {et~al.}, ``{PDF4LHC recommendations for LHC Run
  II}'',} \textit{ J. Phys. G} \textbf{ 43} (2016) 023001,
  \href{http://dx.doi.org/10.1088/0954-3899/43/2/023001}{\doi{10.1088/0954-3899/43/2/023001}},
\href{http://www.arXiv.org/abs/1510.03865}{\texttt{arXiv:1510.03865}}.

\bibitem{Bourilkov:2016qum}
\hrefCMSnoop {}{D.~Bourilkov, ``{Photon-induced background for dilepton
  searches and measurements in pp collisions at 13 TeV}'',} (2016).
\href{http://www.arXiv.org/abs/1606.00523}{\texttt{arXiv:1606.00523}}.

\bibitem{Bourilkov:2016oet}
\hrefCMSnoop {}{D.~Bourilkov, ``{Exploring the LHC Landscape with
  dileptons}'',} (2016).
\href{http://www.arXiv.org/abs/1609.08994}{\texttt{arXiv:1609.08994}}.

\bibitem{Czakon:2011xx}
\hrefCMSnoop {}{M.~Czakon and A.~Mitov, ``{Top++: a program for the calculation
  of the top-pair cross-section at hadron colliders}'',} \textit{ Comput. Phys.
  Commun.} \textbf{ 185} (2014) 2930,
  \href{http://dx.doi.org/10.1016/j.cpc.2014.06.021}{\doi{10.1016/j.cpc.2014.06.021}},
\href{http://www.arXiv.org/abs/1112.5675}{\texttt{arXiv:1112.5675}}.

\bibitem{Alwall:2014hca}
J.~Alwall\hrefCMSnoop {}{ {et~al.}, ``The automated computation of tree-level
  and next-to-leading order differential cross sections and their matching to
  parton shower simulations'',} \textit{ JHEP} \textbf{ 07} (2014) 079,
  \href{http://dx.doi.org/10.1007/JHEP07(2014)079}{\doi{10.1007/JHEP07(2014)079}},
  \href{http://www.arXiv.org/abs/1405.0301}{\texttt{arXiv:1405.0301}}.

\bibitem{Whalley:2005nh}
\hrefCMSnoop {}{M.~R. Whalley, D.~Bourilkov, and R.~C. Group, ``{The Les
  Houches accord PDFs (LHAPDF) and LHAGLUE}'',} in \textit{ {HERA and the LHC:
  A Workshop on the implications of HERA for LHC physics. Proceedings, Part
  B}}.
\newblock 2005.
\newblock
\href{http://www.arXiv.org/abs/hep-ph/0508110}{\texttt{arXiv:hep-ph/0508110}}.
\newblock

\bibitem{Bourilkov:2006cj}
\hrefCMSnoop {}{D.~Bourilkov, R.~C. Group, and M.~R. Whalley, ``{LHAPDF: PDF
  use from the Tevatron to the LHC}'',} in \textit{ {TeV4LHC Workshop - 4th
  meeting Batavia, Illinois, October 20-22, 2005}}.
\newblock (2006).
\newblock
\href{http://www.arXiv.org/abs/hep-ph/0605240}{\texttt{arXiv:hep-ph/0605240}}.
\newblock

\bibitem{Buckley:2014ana}
A.~Buckley\hrefCMSnoop {}{ {et~al.}, ``{LHAPDF6: parton density access in the
  LHC precision era}'',} \textit{ Eur. Phys. J. C} \textbf{ 75} (2015) 132,
  \href{http://dx.doi.org/10.1140/epjc/s10052-015-3318-8}{\doi{10.1140/epjc/s10052-015-3318-8}},
\href{http://www.arXiv.org/abs/1412.7420}{\texttt{arXiv:1412.7420}}.

\bibitem{Agostinelli:2002hh}
\hrefCMSnoop {}{{GEANT4} Collaboration, ``{{\GEANTfour}---a simulation
  toolkit}'',} \textit{ Nucl. Instrum. Meth. A} \textbf{ 506} (2003) 250,
\href{http://dx.doi.org/10.1016/S0168-9002(03)01368-8}{\doi{10.1016/S0168-9002(03)01368-8}}.

\bibitem{CMS-ele-paper}
\hrefCMSnoop {}{{CMS Collaboration}, ``Performance of electron reconstruction
  and selection with the {CMS} detector in proton-proton collisions at
  {$\sqrt{s}=8$ TeV}'',} \textit{ JINST} \textbf{ 10} (2015) P06005,
  \href{http://dx.doi.org/10.1088/1748-0221/10/06/P06005}{\doi{10.1088/1748-0221/10/06/P06005}},
\href{http://www.arXiv.org/abs/1502.02701}{\texttt{arXiv:1502.02701}}.

\bibitem{MUO-10-004-PAS}
\hrefCMSnoop {}{{CMS Collaboration}, ``Performance of {CMS} muon reconstruction
  in {$\Pp\Pp$} collision events at $\sqrt{s} = 7$ {TeV}'',} \textit{ JINST}
  \textbf{ 7} (2012) P10002,
  \href{http://dx.doi.org/10.1088/1748-0221/7/10/P10002}{\doi{10.1088/1748-0221/7/10/P10002}},
\href{http://www.arXiv.org/abs/1206.4071}{\texttt{arXiv:1206.4071}}.

\bibitem{CMS-NOTE-2011-005}
\href {http://cdsweb.cern.ch/record/1379837}{{ATLAS and CMS Collaborations},
  ``Procedure for the {LHC} {H}iggs boson search combination in {S}ummer
  2011'',} Technical Report CMS-NOTE-2011-005, ATL-PHYS-PUB-2011-11, 2011.

\bibitem{Moneta:2010pm}
L.~Moneta\href
  {http://pos.sissa.it/archive/conferences/093/057/ACAT2010_057.pdf}{ {et~al.},
  ``The {R}oo{S}tats {P}roject'',} in \textit{ 13$^\text{th}$ International
  Workshop on Advanced Computing and Analysis Techniques in Physics Research
  (ACAT2010)}.
\newblock SISSA, 2010.
\newblock \href{http://www.arXiv.org/abs/1009.1003}{\texttt{arXiv:1009.1003}}.
\newblock {PoS(ACAT2010)057}.

\bibitem{delAmoSanchez:2010ae}
\hrefCMSnoop {}{{BaBar} Collaboration, ``{Study of $B \to X\gamma$ decays and
  determination of $|V_{td}/V_{ts}|$}'',} \textit{ Phys. Rev. D} \textbf{ 82}
  (2010) 051101,
  \href{http://dx.doi.org/10.1103/PhysRevD.82.051101}{\doi{10.1103/PhysRevD.82.051101}},
\href{http://www.arXiv.org/abs/1005.4087}{\texttt{arXiv:1005.4087}}.

\bibitem{Oreglia:1980cs}
\href {http://www.slac.stanford.edu/cgi-wrap/getdoc/slac-r-236.pdf}{M.~J.
  Oreglia, ``A study of the reactions $\psi^\prime \to \gamma \gamma \psi$''}.
\newblock PhD thesis, {Stanford University}, 1980.
\newblock {SLAC} Report {SLAC-R-236}.

\bibitem{Ball:2012cx}
\hrefCMSnoop {}{{NNPDF} Collaboration, ``{Parton distributions with LHC
  data}'',} \textit{ Nucl. Phys. B} \textbf{ 867} (2013) 244,
  \href{http://dx.doi.org/10.1016/j.nuclphysb.2012.10.003}{\doi{10.1016/j.nuclphysb.2012.10.003}},
\href{http://www.arXiv.org/abs/1207.1303}{\texttt{arXiv:1207.1303}}.

\bibitem{Gross:2010qma}
\hrefCMSnoop {}{E.~Gross and O.~Vitells, ``{Trial factors or the look elsewhere
  effect in high energy physics}'',} \textit{ Eur. Phys. J. C} \textbf{ 70}
  (2010) 525,
  \href{http://dx.doi.org/10.1140/epjc/s10052-010-1470-8}{\doi{10.1140/epjc/s10052-010-1470-8}},
\href{http://www.arXiv.org/abs/1005.1891}{\texttt{arXiv:1005.1891}}.

\bibitem{Chatrchyan:2012xdj}
\hrefCMSnoop {}{{CMS Collaboration}, ``{Observation of a new boson at a mass of
  125 GeV with the CMS experiment at the LHC}'',} \textit{ Phys. Lett. B}
  \textbf{ 716} (2012) 30,
  \href{http://dx.doi.org/10.1016/j.physletb.2012.08.021}{\doi{10.1016/j.physletb.2012.08.021}},
\href{http://www.arXiv.org/abs/1207.7235}{\texttt{arXiv:1207.7235}}.

\bibitem{Mathews:2005bw}
\hrefCMSnoop {}{P.~Mathews, V.~Ravindran, and K.~Sridhar, ``{NLO-QCD}
  corrections to dilepton production in the {R}andall--{S}undrum model'',}
  \textit{ JHEP} \textbf{ 10} (2005) 031,
  \href{http://dx.doi.org/10.1088/1126-6708/2005/10/031}{\doi{10.1088/1126-6708/2005/10/031}},
\href{http://www.arXiv.org/abs/hep-ph/0506158}{\texttt{arXiv:hep-ph/0506158}}.

\bibitem{Alwall:2014bza}
J.~Alwall\hrefCMSnoop {}{ {et~al.}, ``{Computing decay rates for new physics
  theories with FeynRules and \MADGRAPH{5}\_a\MCATNLO}'',} \textit{ Comput.
  Phys. Commun.} \textbf{ 197} (2015) 312,
  \href{http://dx.doi.org/10.1016/j.cpc.2015.08.031}{\doi{10.1016/j.cpc.2015.08.031}},
\href{http://www.arXiv.org/abs/1402.1178}{\texttt{arXiv:1402.1178}}.

\bibitem{Backovic:2015tpt}
M.~Backovi\'{c}\hrefCMSnoop {}{ {et~al.}, ``{MadDM: New dark matter tool in the
  LHC era}'',} \textit{ AIP Conf. Proc.} \textbf{ 1743} (2016) 060001,
  \href{http://dx.doi.org/10.1063/1.4953318}{\doi{10.1063/1.4953318}},
\href{http://www.arXiv.org/abs/1509.03683}{\texttt{arXiv:1509.03683}}.

\bibitem{Backovic:2013dpa}
\hrefCMSnoop {}{M.~Backovi\'{c}, K.~Kong, and M.~McCaskey, ``{MadDM v.1.0:
  Computation of Dark Matter Relic Abundance using \MADGRAPH{5}}'',} \textit{
  Phys. Dark Univ.} \textbf{ 5-6} (2014) 18,
  \href{http://dx.doi.org/10.1016/j.dark.2014.04.001}{\doi{10.1016/j.dark.2014.04.001}},
\href{http://www.arXiv.org/abs/1308.4955}{\texttt{arXiv:1308.4955}}.

\bibitem{Sirunyan:2017hci}
\hrefCMSnoop {}{{CMS Collaboration}, ``{Search for dark matter produced with an
  energetic jet or a hadronically decaying W or Z boson at $ \sqrt{s}=13 $
  TeV}'',} \textit{ JHEP} \textbf{ 07} (2017) 014,
  \href{http://dx.doi.org/10.1007/JHEP07(2017)014}{\doi{10.1007/JHEP07(2017)014}},
\href{http://www.arXiv.org/abs/1703.01651}{\texttt{arXiv:1703.01651}}.

\end{thebibliography}\endgroup
\cleardoublepage \appendix\section{The CMS Collaboration \label{app:collab}}\begin{sloppypar}\hyphenpenalty=5000\widowpenalty=500\clubpenalty=5000\vskip\cmsinstskip
\textbf{Yerevan Physics Institute,  Yerevan,  Armenia}\\*[0pt]
A.M.~Sirunyan,  A.~Tumasyan
\vskip\cmsinstskip
\textbf{Institut f\"{u}r Hochenergiephysik,  Wien,  Austria}\\*[0pt]
W.~Adam,  F.~Ambrogi,  E.~Asilar,  T.~Bergauer,  J.~Brandstetter,  E.~Brondolin,  M.~Dragicevic,  J.~Er\"{o},  A.~Escalante Del Valle,  M.~Flechl,  M.~Friedl,  R.~Fr\"{u}hwirth\cmsAuthorMark{1},  V.M.~Ghete,  J.~Grossmann,  J.~Hrubec,  M.~Jeitler\cmsAuthorMark{1},  A.~K\"{o}nig,  N.~Krammer,  I.~Kr\"{a}tschmer,  D.~Liko,  T.~Madlener,  I.~Mikulec,  E.~Pree,  N.~Rad,  H.~Rohringer,  J.~Schieck\cmsAuthorMark{1},  R.~Sch\"{o}fbeck,  M.~Spanring,  D.~Spitzbart,  A.~Taurok,  W.~Waltenberger,  J.~Wittmann,  C.-E.~Wulz\cmsAuthorMark{1},  M.~Zarucki
% ...
\vskip\cmsinstskip
% ...
1:~Also at Vienna University of Technology,  Vienna,  Austria\\
% ...
\end{sloppypar}
\end{document}